\documentclass{aa}
\usepackage{graphicx}
\usepackage{multirow}
\usepackage{txfonts}
\usepackage[]{hyperref}

\begin{document} 

   \title{Observability of hydrogen-rich exospheres in Earth-like exoplanets}

   \author{Leonardo A. dos Santos
          \inst{1}
          \and
          Vincent Bourrier\inst{1}
          \and
          David Ehrenreich\inst{1}
          \and
          Shingo Kameda\inst{2}
          }

   \institute{Observatoire astronomique de l’Université de Genève, 51 
   			  chemin des Maillettes, 1290 Versoix, Switzerland\\
              \email{Leonardo.dosSantos@unige.ch}
         \and
          Department of Physics, College of Science, Rikkyo University,
          3 Chome-34-1 Nishiikebukuro, Tokyo 171-8501, Japan \\
             }

   \date{Received 08 May, 2018; accepted 28 November, 2018}

  \abstract
   {The existence of an extended neutral hydrogen exosphere around small planets can be used as an evidence for the presence of water in their lower atmosphere but, to date, such feature has not been securely detected in rocky exoplanets. Planetary exospheres can be observed using transit spectroscopy of the Lyman-$\alpha$ line, which is limited mainly by interstellar medium absorption in the core of the line, and airglow contamination from the geocorona when using low-orbit space telescopes.}
   {Our objective is to assess the detectability of the neutral hydrogen exosphere of an Earth-like planet transiting a nearby M dwarf using Lyman-$\alpha$ spectroscopy and provide the necessary strategies to inform future observations.}
   {Our tests require spatial and velocity information of the neutral hydrogen particles in the upper atmosphere. The spatial distribution is provided by an empirical model of the geocorona, and we assume a velocity distribution based on radiative pressure as the main driver in shaping the exosphere. We compute the excess absorption in the stellar Lyman-$\alpha$ line while in transit, and use realistic estimates of the uncertainties involved in observations to determine the observability of the signal.}
   {We found that the signal in Lyman-$\alpha$ of the exosphere of an Earth-like exoplanet transiting M dwarfs with radii between 0.1 and 0.6 R$_\odot$ produces an excess absorption between 50 and 600 ppm. The Lyman-$\alpha$ flux of stars decays exponentially with distance because of interstellar medium absorption, which is the main observability limitation. Other limits are related to the stellar radial velocity and instrumental setup.}
   {The excess absorption in Lyman-$\alpha$ is observable using LUVOIR/LUMOS in M dwarfs up to a distance of $\sim$15 pc. The analysis of noise-injected data suggests that it would be possible to detect the exosphere of an Earth-like planet transiting TRAPPIST-1 within 20 transits.}


   \titlerunning{Observability of exospheres in Earth-like exoplanets}
   \maketitle
%

\section{Introduction}

\defcitealias{2013A&A...557A.124B}{BL13}

The exosphere of a planet is defined as the region where the gas becomes collisionless, or more formally the region where the Knudsen number (the ratio between the mean free path and the scale height) is higher than one. This region is predominantly sustained by atmospheric escape from the planet, driven mainly by non-thermal processes \citep{1996RvGeo..34..483S, 2011ApJ...729L..24V} for planets like the Earth, and thermal hydrodynamic escape for strongly irradiated giant planets \citep[e.g.,][]{2003ApJ...598L.121L, 2004A&A...418L...1L}.

The first observations of evaporating atmospheres of exoplanets were performed for hot Jupiters such as HD~209458~b \citep{2003Natur.422..143V, 2008A&A...483..933E, 2010ApJ...717.1291L} and HD~189733~b \citep{2010A&A...514A..72L, 2012A&A...543L...4L, 2013ApJ...773...62P}. The process of evaporation is facilitated in such close-in inflated planets because, in the energy-limited regime, the mass loss rate is proportional to the high-energy flux received from the host star and inversely proportional to the bulk density of the planet \citep{2007A&A...461.1185L}. Atmospheric escape has also been observed in the warm Neptune exoplanet GJ~436~b \citep{2014ApJ...786..132K, 2015Natur.522..459E}, in which the conditions are favorable to produce a comet-like tail of neutral hydrogen \citep[][]{2015A&A...582A..65B, 2016A&A...591A.121B, 2017A&A...605L...7L}.

An exospheric cloud such as the one around GJ~436~b produces extremely large transit absorption signatures in Lyman-$\alpha$ \citep[56\% $\pm$ 3\%,][]{2015Natur.522..459E} compared to what is measured in the optical transmission spectrum of this planet \citep[250 ppm,][]{2018AJ....155...66L}. It is not known how small a planet can be and still exhibit a readily detectable exospheric cloud. The Earth itself is surrounded by a neutral hydrogen exosphere, which produces the geocorona\footnote{\footnotesize{The traditional definition of geocorona is the light emitted by the Earth's exosphere. In this manuscript, however, we use the terms "Earth's exosphere" and "geocorona" interchangeably.}} \citep{1970JGR....75.3769W, 1972Sci...177..788C}. A recent Lyman-$\alpha$ (hereafter Ly$\alpha$) image of the Earth taken from 0.1 au showed that the geocorona spans more than 38 Earth radii \citep{2017GeoRL..4411706K}. The hydrogen density is approximately 20 atoms~cm$^{-3}$ at a height of 600,000 km in the Earth's exosphere. Venus and Mars also have H-rich exospheres, but they are far less extended and less dense because these planets have atmospheres rich in CO$_2$ and possess lower exospheric temperatures \citep{2007SSRv..129..207K}.

Detecting an extended hydrogen exosphere around a rocky exoplanet would be a compelling evidence for the presence of water in the lower atmosphere of the planet, since the atomic H is a product of photodissociation of water. \citet{2004ApJ...605L..65J} indeed proposed that evaporating exoplanetary oceans could fuel H-rich exospheres \citep[see also][]{2007Icar..191..453S, 2017MNRAS.464.3728B}. This could be a new path to characterize habitability-zone (HZ) exoplanets and trace their water content\footnote{\footnotesize{The direct detection of water in the atmosphere of water-rich planets will be challenging, even with future infrared spectrographs such as JWST \citep[see, e.g.,][]{2017JGRE..122...53D}.}}. This is especially relevant for planets around M dwarfs, which are targets of intense search for rocky planets because their HZ is closer to the star and transiting planets produce large signals. Much speculation has been done, for instance, for the water content of the TRAPPIST-1 planets \citep{2016Natur.533..221G, 2017A&A...599L...3B, 2017AJ....154..121B, 2017MNRAS.464.3728B}.

Far-extended exospheres are easier to analyze in Ly$\alpha$ than the compact lower atmospheres at longer wavelengths. Moreover, the velocity field in the exosphere can reach more than a hundred km s$^{-1}$, spreading the atmospheric absorption signal over a large range of wavelengths and increasing its detectability. Currently, the only telescope that has access to the Ly$\alpha$ line is the HST, and the instrument usually employed for these observations is the Space Telescope Imaging Spectrograph (STIS). In the future, projects such as the Large Ultraviolet Optical Infrared telescope \citep[LUVOIR,][]{2017SPIE10398E..09B} and the Habitable Planet Explorer \citep[HabEx,][]{2018NatAs...2..600G} will build upon on the FUV capabilities of HST in several aspects. Using a theoretical model of the Earth's exosphere and several simplifications, \citet{2018ExA...tmp....8C} calculated that the detection of an Earth-like exosphere is possible for the nearest stars, provided that the mirror size is larger than 4 m. In this work, we perform a more detailed and extensive assessment of this prospect based on observational constraints of the geocorona \citep{2017GeoRL..4411706K} and provide scaling relations for the stellar radius, distance (considering interstellar medium absorption), exospheric density and instrumental setup. Further, this study aims at informing the community about the capabilities of future UV telescopes to characterize HZ exoplanets.

This manuscript has the following structure: in Sect. \ref{obs_mod} we present the observation and models of the Earth's exosphere; in Sect. \ref{lya-m-dwarfs} we describe how the Ly$\alpha$ profile of different targets and conditions is computed and estimate the excess absorption signal in Ly$\alpha$ transmission spectroscopy for an Earth-like planet transiting a M dwarf; in Sect. \ref{obs_strategy} we extensively discuss the observation strategies and limitation on performing such a detection; in Sect. \ref{toi} we briefly present the best targets on which to conduct intensive searches; in Sect. \ref{conclusions} we present the main conclusions obtained in this research.

\section{Observation and model of the Earth's hydrogen exosphere}\label{obs_mod}

\subsection{The neutral hydrogen geocorona}

The Earth's exosphere was observed using the Lyman Alpha Imaging Camera (LAICA) onboard the Proximate Object Close Flyby with Optical Navigation (PROCYON) on 9 January 2015, when the spacecraft was at a distance of 2348~R$_\oplus$ ($0.1$~au) from the Earth and at coordinates 74.09$^\circ$ and -23.94$^\circ$ in right ascension and declination \citep[see figure 1 in][]{2017GeoRL..4411706K}. The LAICA instrumentation consists of a 41-mm diameter Cassegrain-type telescope and a detector unit equipped with a $122 \pm 10$ nm band-pass wavelength filter. The total exposure time of the observation was 300 s, spatial resolution was 1.34 and 0.98 R$_\oplus$/pixel in the horizontal and vertical directions, respectively, and the effective field of view of LAICA is $2^\circ \times 2^\circ$. This observation revealed that the geocorona spans more than 38 R$_\oplus$, a value that exceeds the previously determined lower limit of 16 R$_\oplus$ \citep{1970JGR....75.3769W, 1972Sci...177..788C}. According to \citet{2017GeoRL..4411706K}, the geocorona possesses north-south symmetry, indicating that magnetic fields are not dominant in shaping the geocorona under moderate solar activity; the asymmetry in the horizontal direction is inconclusive, since it could be simply explained by the viewing geometry of the observations. These observations did not include high-resolution spectroscopy, thus they do not provide direct dynamical information about the geocorona.

Based on this observation, \citet{2017GeoRL..4411706K} reconstructed the geocorona using a modified Chamberlain model \citep{1963P&SS...11..901C} with three parameters: the exobase temperature, exobase density and solar radiation pressure. The force of the solar radiation was determined using the local photon scattering rate; assuming an exobase temperature of 1000 K \citep[as in][]{2002JGRA..107.1468P}, the exobase density was derived from the fit to the emission data using a simple $\chi^2$ minimization.

\subsection{Densities and velocity field}\label{dens}

\begin{figure}
  \centering
  \includegraphics[width=\hsize]{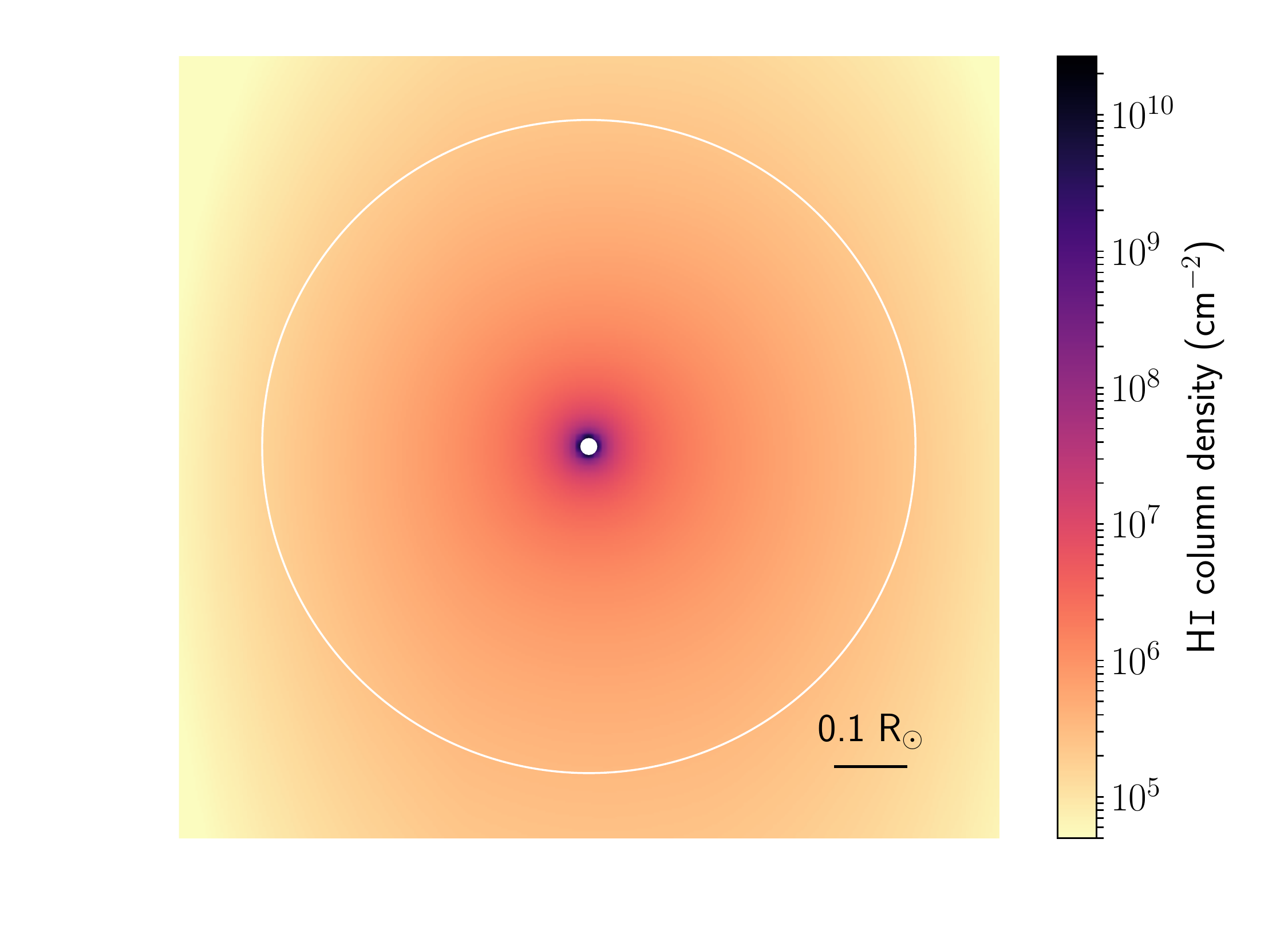}
     \caption{Column density map of the model of the geocorona obtained by \citet{2017GeoRL..4411706K}. The Earth's transiting disk is show as the small white circle. The large white circle represents the size of the M dwarf GJ~436. Asymmetries near the borders are artificial due to rotation of a box-shaped model.}
        \label{GJ-transit}
  \end{figure}

To verify if an Earth-like exosphere around a rocky extra-solar planet is observable with current instrumentation in transit at Ly$\alpha$, we start from the analytical model of the neutral hydrogen geocorona produced by \citet{2017GeoRL..4411706K}. We calculated the three-dimensional density distribution using the model and computed the column densities by integrating the volumetric densities in the line of sight (see Fig. \ref{GJ-transit}).

The geocorona model yields the densities of neutral hydrogen, but not their velocities distribution, which is necessary to compute the Ly$\alpha$ absorption profile. In order to estimate this distribution, we assume that there are two competing forces dictating the dynamics of the neutral hydrogen particles in the exosphere: gravity and radiation pressure. The second can either overturn stellar gravity and move particles away from the star \citep["radiative blow-out,"][]{2013A&A...557A.124B} or slow down the infall of matter towards the star \citep["radiative braking"; see the case of GJ~436~b, a moderately irradiated warm Neptune in figs. 1 and 2 in][]{2015A&A...582A..65B}. Effectively, the line-of-sight velocities of the particles will be limited in the bluer end by the orbital velocity of the planet and in the redder end by the mode of a thermal distribution of particles with a kinetic temperature of 1000 K \citep[mean temperature at the exobase;][]{2017GeoRL..4411706K}, since the line-of-sight component of the orbital velocity at transit center is null for circular orbits. Using this temperature, we obtain that the upper limit of velocities of particles in Earth's exosphere is approximately 4 km s$^{-1}$.

In the reference frame of the host star, particles escaping from an exoplanet atmosphere have an initial velocity that combines the orbital velocity of the planet and the velocity of the planetary wind. Assuming nearly circular orbits, the orbital speed of a planet in relation to the host star is 

\begin{equation}
  v_\mathrm{orb} = \sqrt{\frac{GM_\star}{a}} \mathrm{,}
\end{equation}where $M_\star$ is the stellar mass and $a$ is the orbital semi-major axis of the planet. Assuming that a planet orbits in the inner edge of the HZ of its host star \citep[similarly to the Earth;][]{2013ApJ...765..131K}, the orbital speeds of the hypothetical planet range from $\sim$30--50 km s$^{-1}$, respectively for $M_\star$ varying from 0.1 to 1 $M_\odot$ (for reference, Earth's orbital speed in relation to the Sun is 29.8 km s$^{-1}$). In addition, the planetary wind of Earth is well inside the Jeans escape regime, which means that particles escaping from the atmosphere are in the high-velocity tail of the Maxwell-Boltzmann distribution. The introduction of other mechanisms such as stellar wind and bow shocks, although they could play a role in determining the distribution of positions and velocities of particles, would add too many free parameters to our model and is not justified at this point. Thus, as a first approximation, and based on the results for the Earth \citep[e.g.,][]{2017GeoRL..4411706K} and evaporating planets \citep[e.g.,][]{2013A&A...557A.124B, 2015A&A...582A..65B}, we consider radiation pressure to be the main driver of the dynamics of neutral hydrogen atoms escaping from Earth-like planets.

\section{The signature of transiting exospheres in Lyman-$\alpha$ emission of quiet M dwarfs}\label{lya-m-dwarfs}

\begin{figure}
  \centering
  \includegraphics[width=\hsize]{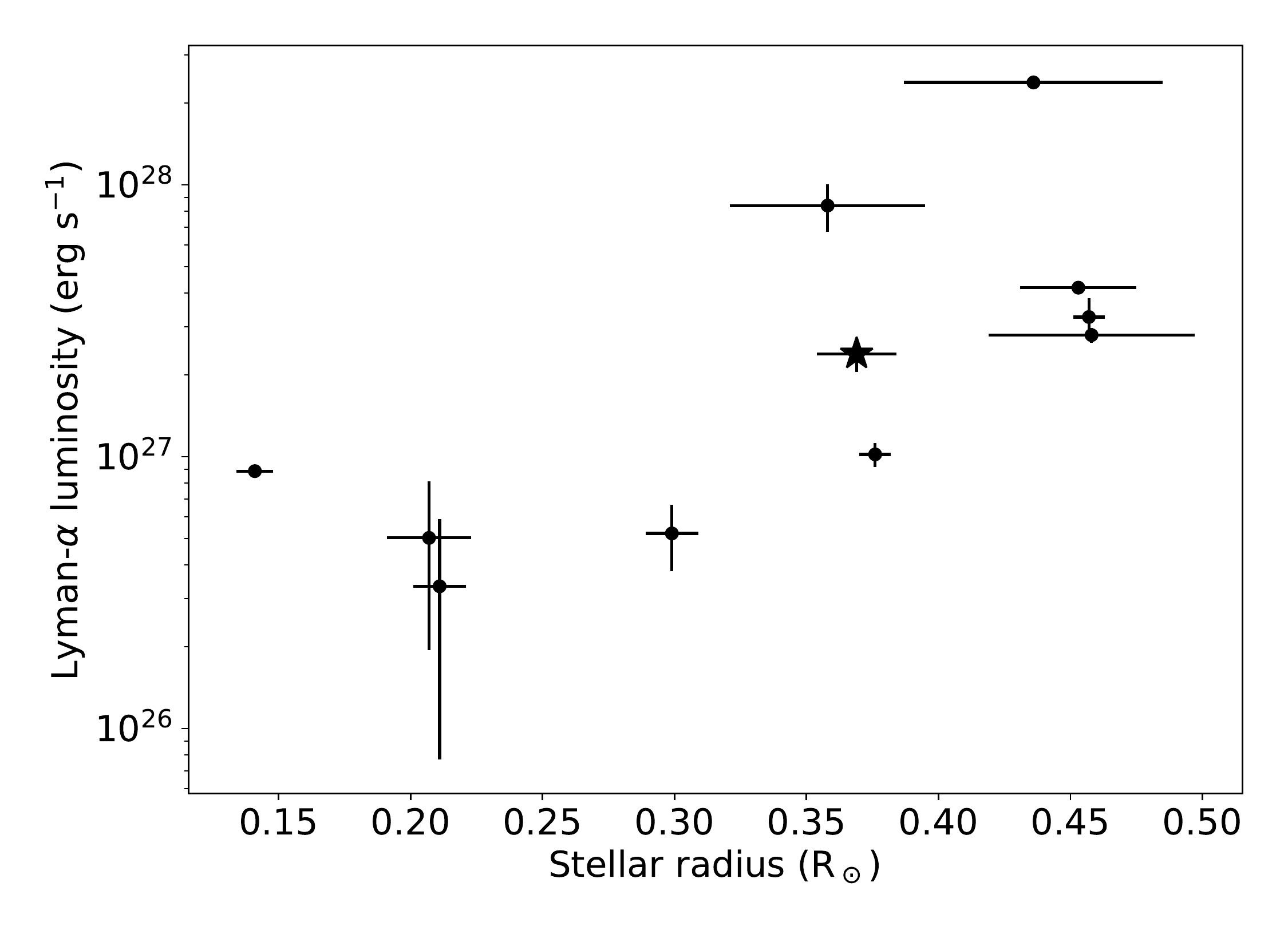}
     \caption{Ly$\alpha$ luminosity of quiet M dwarfs in the MUSCLES sample (based on GAIA DR2 distances). The Ly$\alpha$ fluxes and stellar radii were reported in \citet{2017ApJ...843...31Y}. The star symbol marks the position of GJ~436.}
        \label{lya_vs_r}
  \end{figure}

In general, the Ly$\alpha$ emission line of M dwarfs possesses a nearly Gaussian shape \citep{2005ApJS..159..118W, 2016ApJ...824..101Y, 2017MNRAS.465L..74W}. The MUSCLES survey obtained the Ly$\alpha$ reconstructed flux for 12 inactive early- to mid-M dwarfs using HST \citep{2017ApJ...843...31Y}, and their results show that the Ly$\alpha$ luminosity among the sample varies by two orders of magnitude and correlate with the rotational period of the star; the Ly$\alpha$ luminosity of GJ~436 is representative of the mean in their sample (see Fig. \ref{lya_vs_r}).

\subsection{Interstellar medium absorption}\label{ism_abs_sect}

Except for the Sun, we do not have access to the intrinsic Ly$\alpha$ emission of stars due to absorption by the neutral hydrogen atoms present in the interstellar medium (ISM) between the target and the Earth. If the ISM clouds in the line of sight have radial velocities close to the radial velocity of the target star, the core of the stellar Ly$\alpha$ emission line is completely absorbed. Thus, in such cases, only the wings of the Ly$\alpha$ line contain information about the system. For distant targets, not only their faintness is a limitation in UV observations, but their distances also increase the column densities of ISM neutral hydrogen in the line of sight, as well as the range of ISM radial velocities; for these reasons, we can only study escaping atmospheres in Ly$\alpha$ for the closest ($d \lesssim 60$ pc) exoplanetary systems, or for host stars with large radial velocity \citep[such as Kepler-444,][]{2017A&A...602A.106B}. There is also absorption by deuterium at the rest wavelength 1215.25 \AA, although less pronounced than the absorption by \ion{H}{I} atoms. The radial velocity of the local interstellar cloud (LIC) varies in the range [-25, 25] km s$^{-1}$, approximately, depending on the direction of sight \citep{2005ApJS..159..118W}.

In order to estimate how the ISM absorption affects the observable Ly$\alpha$ fluxes of M dwarfs with different distances, we simulated the total attenuation of the flux caused by ISM absorption. We set the physical properties of the ISM to be isotropic, namely the temperature to 8000 K (similar to the what is measured in the solar neighborhood), the radial velocity to zero and turbulence velocity to zero (other values of turbulence do not significantly affect our results and conclusions). We assumed the minimum and maximum densities of neutral H in the solar neighborhood (0.01 and 0.1 cm$^{-3}$), and adopted the fraction between the densities of deuterium and H $N_\mathrm{D} / N_\mathrm{H} = 1.5 \times 10^{-5}$ \citep{2013A&A...557A.124B}. The ISM absorption profiles are computed as in \citet{2017A&A...597A..26B}.

\begin{figure*}
  \centering
  \begin{tabular}{cc}
  \includegraphics[width=0.47\hsize]{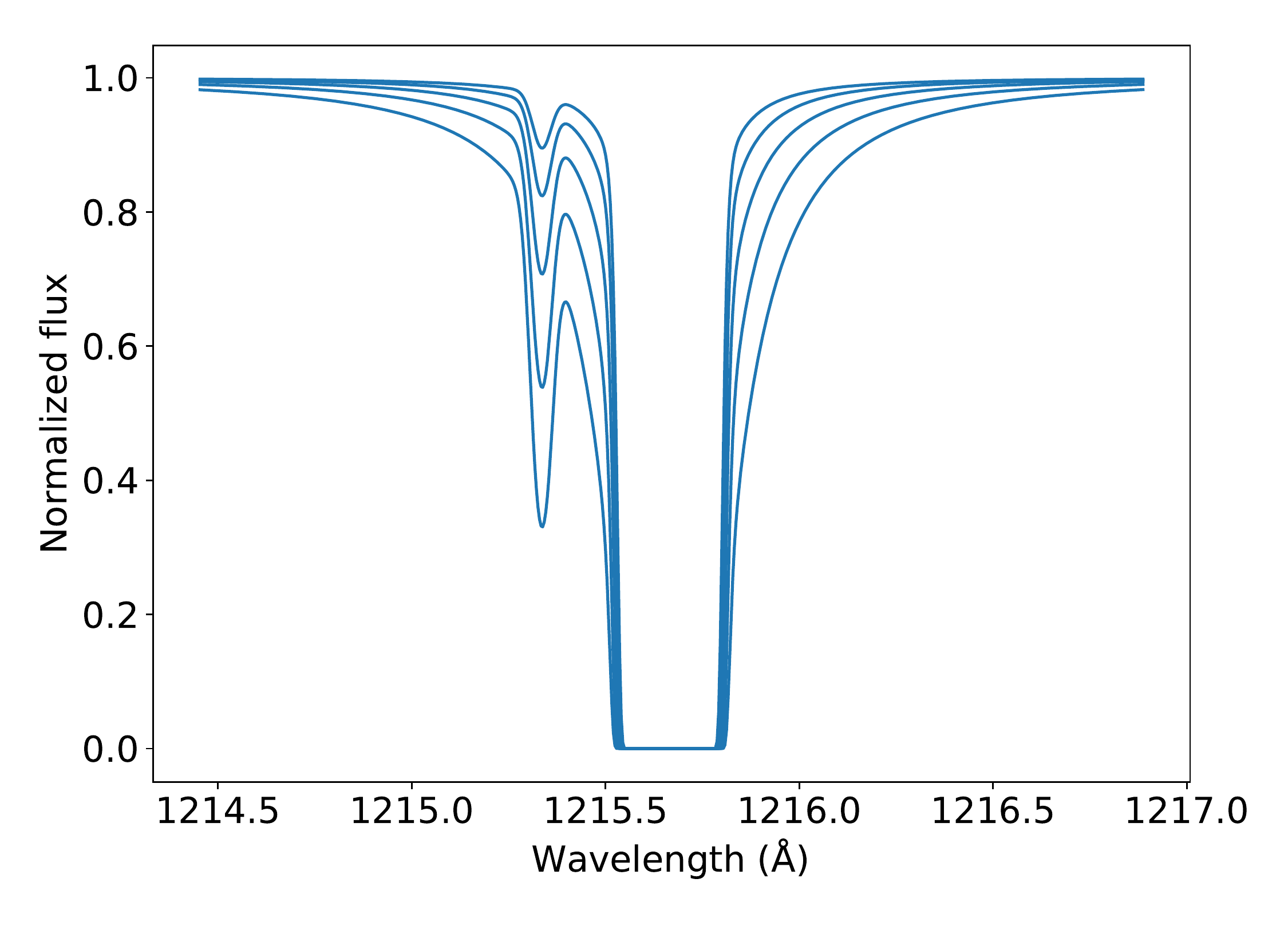} & \includegraphics[width=0.47\hsize]{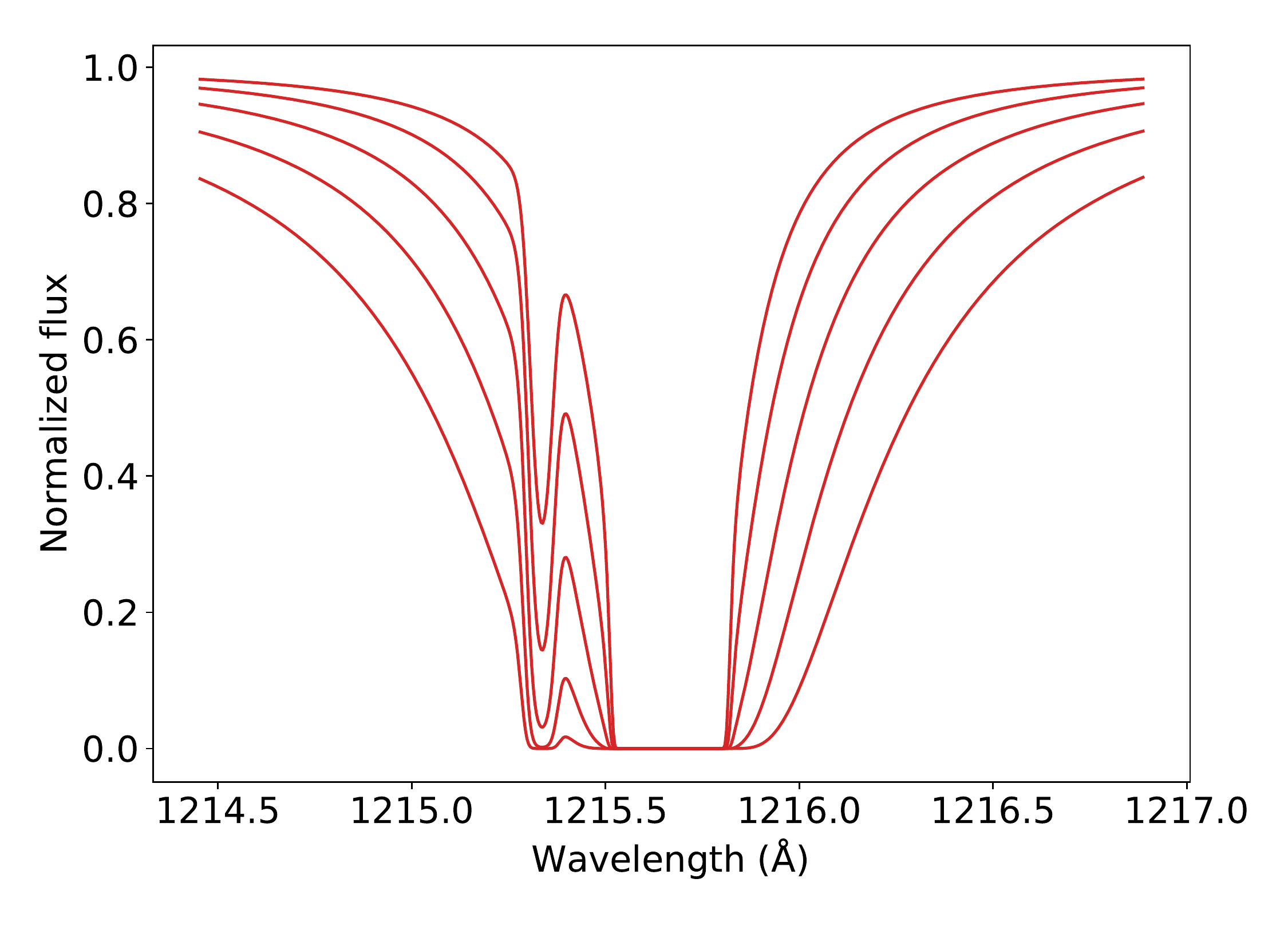} \\
  \end{tabular}
     \caption{Left panel: Expected interstellar medium absorption profiles for distances 2.0, 3.6, 6.3, 11.2 and 20.0 pc and for the minimum column density of the local interstellar cloud (radial velocity of the cloud is fixed at zero). Right panel: Same as left panel, but for the maximum column density of the LIC.}
        \label{ism_abs_profiles}
  \end{figure*}

We found that the Ly$\alpha$ line fluxes of M dwarfs decrease approximately exponentially with the distance in the range from 2 to 20 pc, where the flux is attenuated to less than 10\% of the intrinsic emission in case we observe in a direction with high H column density. However, we reiterate that this is an approximation, since it is well-known that the ISM in the solar neighborhood is patchy \citep{2008ApJ...673..283R} and its physical properties vary, so in reality the attenuation will depend on the direction of observation.

\begin{figure}
  \centering
  \includegraphics[width=\hsize]{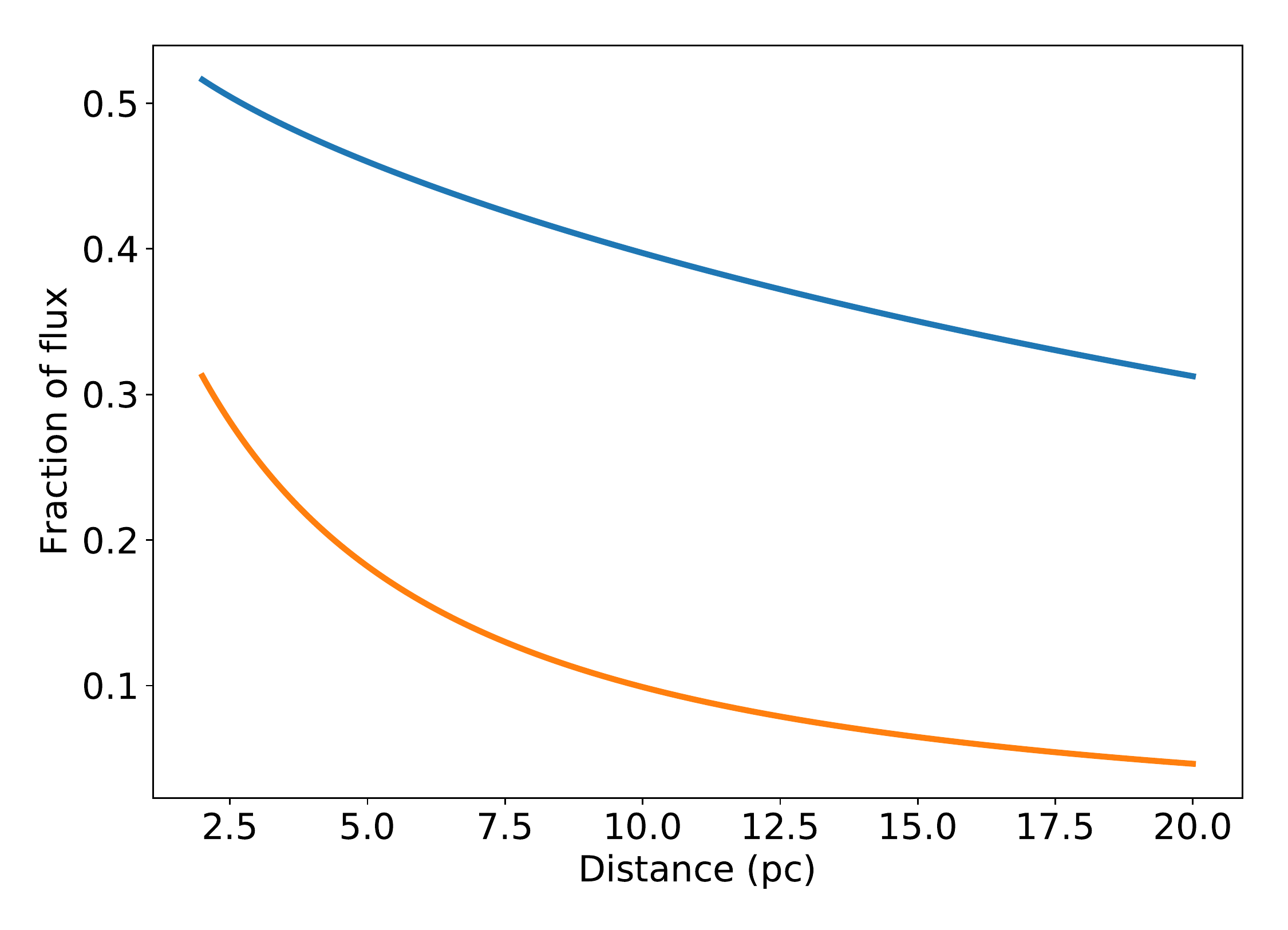}
     \caption{Observed fraction of the intrinsic Lyman-$\alpha$ flux in function of distance after ISM absorption. The blue (orange) curve represents the attenuation in the lowest (highest) density of the LIC.}
        \label{ism_att}
  \end{figure}

\subsection{Estimating the excess absorption}

When a planet with a H-rich exosphere transits its host star, the opaque disk of the planet will block part of the Ly$\alpha$ flux (i.e. baseline continuum absorption) and its exosphere will absorb part of the flux in addition (i.e. excess absorption).

The absorption profile caused by the passage of a cloud of neutral hydrogen in front of a stellar disk is computed following the equations in Sect. 2.4 of \citet{2013A&A...557A.124B} applied to the model proposed by \citet{2017GeoRL..4411706K}. The absorption profile depends mainly on the spatial and velocity distributions of neutral hydrogen atoms (see Sect. \ref{dens}). In order to assess the excess absorption by a transiting exosphere, we use synthetic Ly$\alpha$ profiles based on the observed emission line of the quiet M dwarf GJ~436.

To obtain the in-transit spectrum, we position an Earth-sized planet and its exosphere in the center of the disk of the star at mid-transit (impact parameter $b = 0$), compute its absorption profile and multiply this profile by the out-of-transit intrinsic stellar spectrum. Both in- and out-of-transit spectra are then absorbed by a realistic ISM model. When measured through a telescope, the observable spectra are convolved with the instrumental response; furthermore, during the analysis we need to avoid using wavelength ranges where the signal-to-noise ratio is too low---we describe practical aspects of observations in Sect. \ref{obs_strategy}.

The exospheric excess absorption per unit wavelength $S_\lambda$ in the Ly$\alpha$ line is measured by the ratio of the in- ($F_{\lambda, \mathrm{in}}$) and out-of-transit flux densities ($F_{\lambda, \mathrm{out}}$). We have, thus

\begin{equation}\label{signal_eq}
S_{\lambda}  = 1 - \frac{F_{\lambda, \mathrm{in}}}{F_{\lambda, \mathrm{out}}} - a_\mathrm{pl} \mathrm{,}
\end{equation}where $a_\mathrm{pl}$ is the baseline absorption caused by the planet's opaque disk. $S_{\lambda}$ does not depend on the distance of the target since the fraction $F_{\lambda, \mathrm{in}} / F_{\lambda, \mathrm{out}}$ cancels out this dependence. The mean excess absorption depth $S$ is measured by taking the mean of $S_\lambda$ between a given range of Doppler velocities.

\begin{figure}
  \centering
  \includegraphics[width=\hsize]{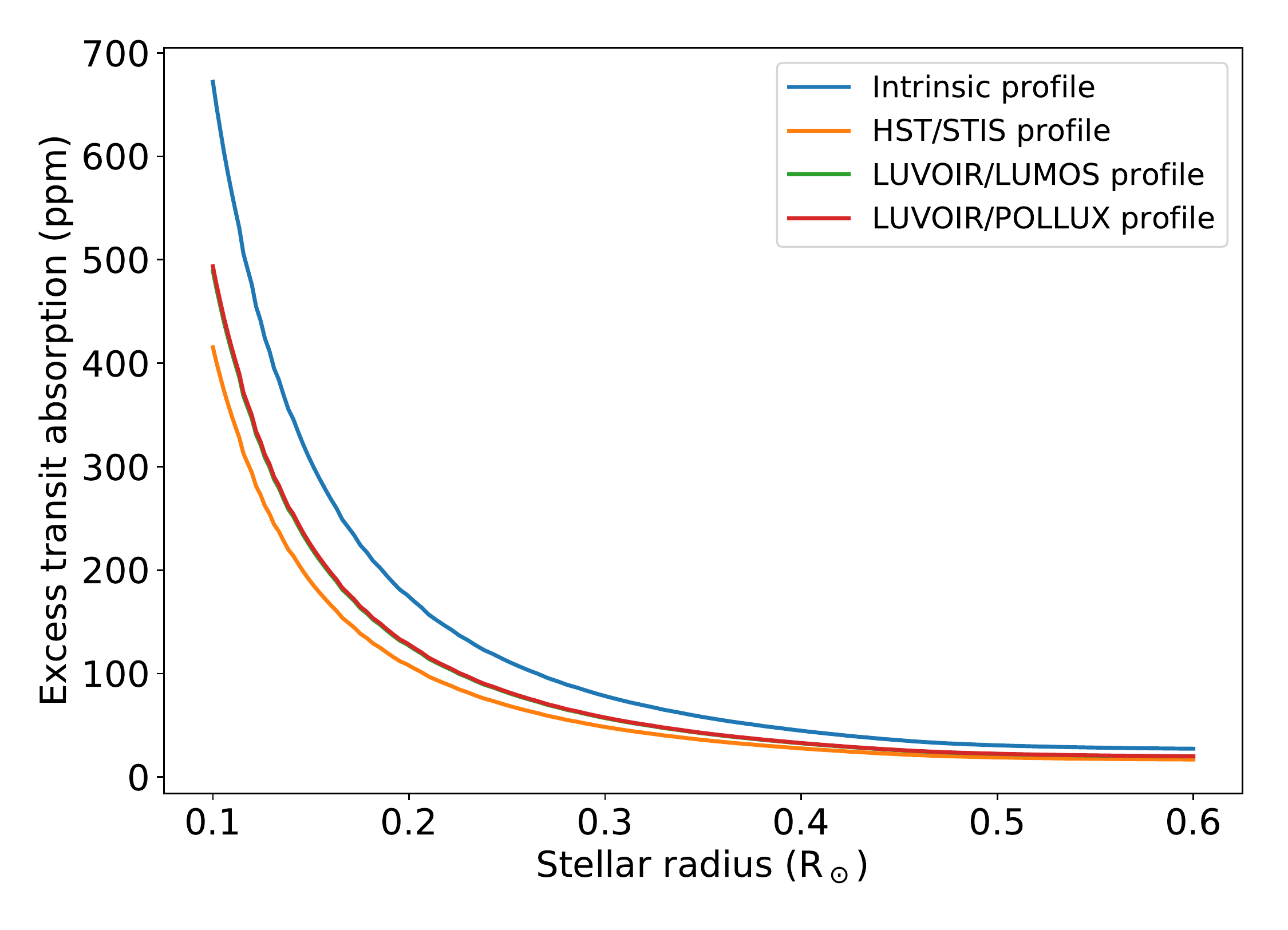}
     \caption{Mean excess transit absorption in Ly$\alpha$ caused by an Earth-like exosphere in function of the stellar radius. In the case of absorption profiles convolved with an instrumental profile, we considered the signal located inside the Doppler velocity range [-50, 10] km s$^{-1}$.}
        \label{t_depth}
  \end{figure}

We simulated a grid of models of transiting events as described above around stars of radii varying between 0.1 and 0.6 R$_\odot$, which is our adopted definition of a M dwarf. Since we assumed a flat distribution of velocities for the particles in the exosphere of the planet, the intrinsic excess transit profile is constant in wavelength. If we take into account the instrumental profile of a spectrograph, this signal is diluted. The resulting relation between stellar size and excess transit absorption is shown in Fig. \ref{t_depth}. Similarly to optical transits, it is clear that excess transit depths are deeper the smaller the host star is, which results from simple geometry.

\begin{figure}
    \centering
    \includegraphics[width=\hsize]{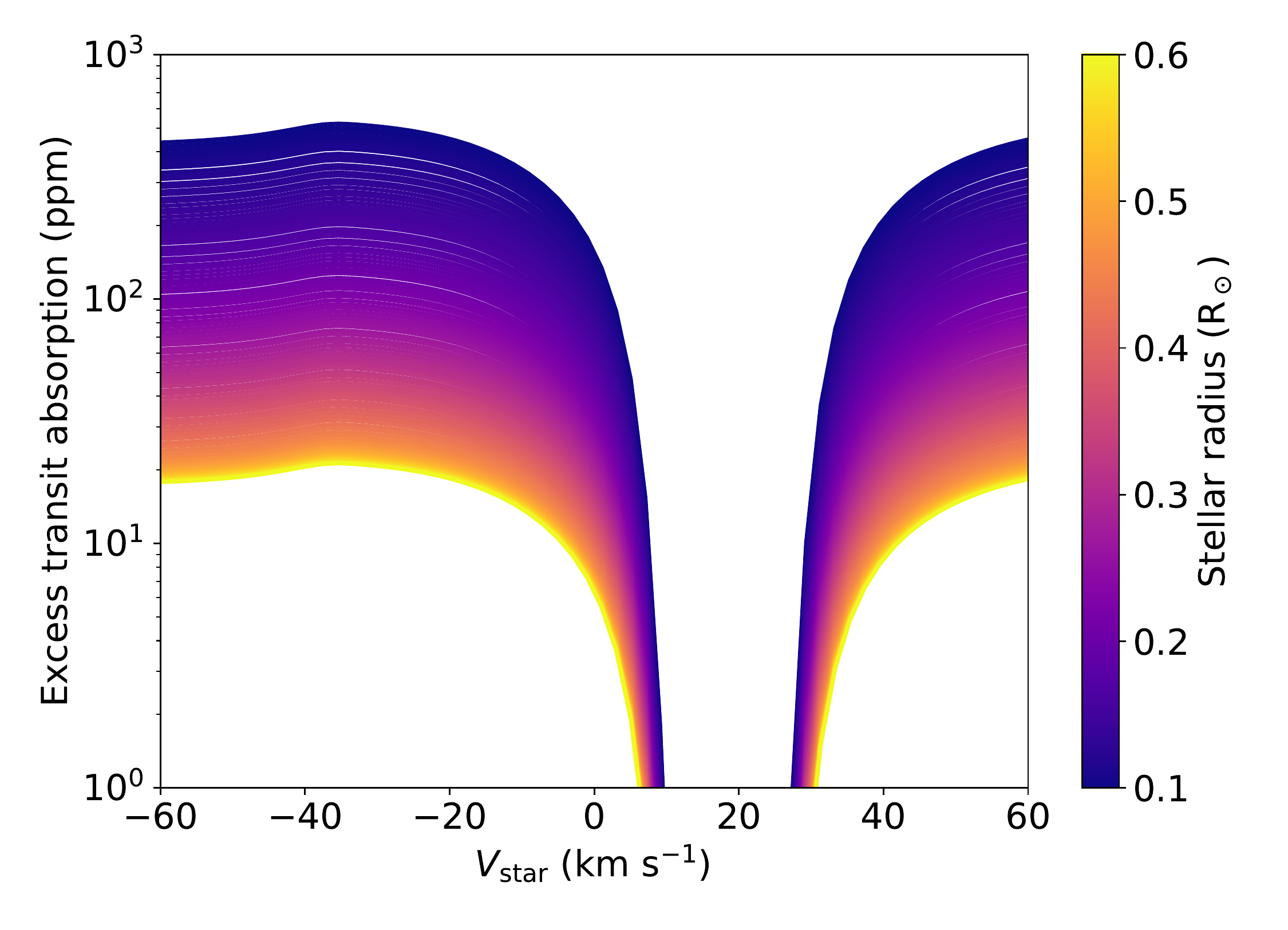}
       \caption{Mean excess transit absorption in Ly$\alpha$ caused by an Earth-like exosphere in function systemic radial velocity as observed with HST/STIS. For spectrographs with higher resolution, the range where the signal is lost is narrower than the range of STIS ($R = 10,000$).}
          \label{signal_hst}
    \end{figure}

We expect $S_{\lambda}$ to depend on the distribution of velocities of H atoms in the exosphere and on the radial velocity of the target star because the ISM absorption quickly saturates with distance and can partially or completely erase the signal we want to observe. Considering our assumptions in Sect. \ref{dens}, the radial velocity range where $S_{\lambda} < 10$ ppm depends mainly on the spectral resolution of the spectrograph: for HST/STIS, this range is located approximately between [0,~40] km s$^{-1}$ (see Fig. \ref{signal_hst}); the approximate ranges for the LUVOIR planned spectrographs LUMOS and POLLUX are, respectively, [5,~30] and [10,~25] km s$^{-1}$. Outside these ranges, the expected signal should follow the relations seen in Fig. \ref{t_depth}.

\section{Practical and observational aspects}\label{obs_strategy}

Currently, the only instruments capable of performing FUV spectroscopy of exoplanetary atmospheres are the STIS and COS spectrographs aboard the HST. Several projects of large-aperture telescopes equipped with FUV instruments are being currently studied. The World Space Observatory \citep[WSO-UV,][]{2016SPIE.9905E..04S}. in particular, is even partially built. In the following sections we are going to consider HST/STIS and two potential instruments with known specifications for the LUVOIR telescope: LUMOS \citep{2016JATIS...2d1203F, 2017SPIE10397E..13F} and POLLUX \citep{2018arXiv180509067M}. We will discuss how these instruments can overcome the current limitations that make it challenging to detect the excess absorption signal of exospheres (see Sect. \ref{lya-m-dwarfs}) around Earth-like planets.

\subsection{Geocoronal contamination}

\begin{figure}
  \centering
  \includegraphics[width=\hsize]{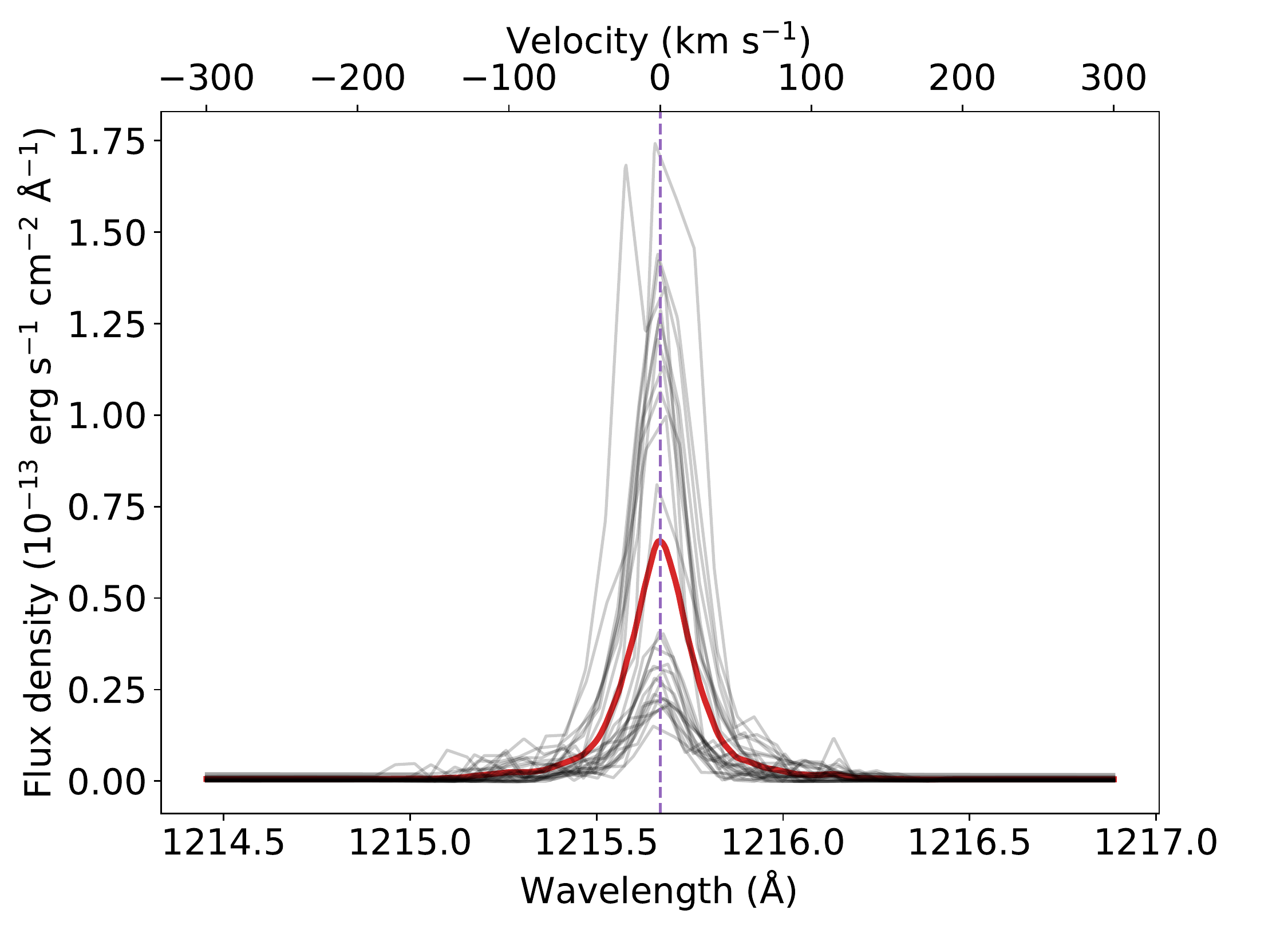}
     \caption{Mean geocoronal emission of the Earth as measured with HST/STIS during 27 observations (red spectrum). The gray spectra correspond to the different exposures. The emission lines were aligned to the rest frame of the telescope by fitting Gaussians to each exposure.}
        \label{geocorona}
  \end{figure}

Far-ultraviolet observations can only be performed from space, and low-orbit telescopes (such as the HST) contain contamination from geocoronal emission produced by the Earth's exosphere (see Fig. \ref{geocorona}). In general, the geocoronal contamination is variable (stronger in the day-side of the Earth and weaker in the night-side), but it can be removed in a straight-forward fashion, provided that the observations are performed with a slit spectrograph: the sky spectrum is measured away from the target, and then it is simply subtracted from the spectrum of the target. According to the manual of HST/STIS, during an exposure with high geocoronal contamination, the fluxes are $\sim$1.98 times larger than the average, and they are $\sim$0.20 times lower when in the shadow of the Earth. In the case of spectrographs with a circular aperture, such as the Cosmic Origins Spectrograph (COS) on HST, the removal of geocoronal contamination is not straight-forward, but it can be performed using airglow templates\footnote{\footnotesize{Airglow templates for HST/COS are available at \url{http://www.stsci.edu/hst/cos/calibration/airglow.html}}} \citep[see, e.g.,][]{2017A&A...599A..75W, 2018A&A...615A.117B}.

In any case, even after the removal of the geocoronal contamination, the additional photon noise will remain in the spectra obtained with low-orbit telescopes. The increased noise will affect how well we can measure the excess absorption signal by the exosphere of a hypothetical Earth-like exoplanet. This issue can be alleviated by observing near the night-side of the Earth, but such a measure will implicate in shorter exposures: in the case of HST, opting for observations in the Earth's shadow will reduce the exposure time by approximately 50\%. It is also more difficult to allocate time for such a proposal that is already time-constrained in the first place.

\subsection{Measurement uncertainties}\label{uncertainties}

The uncertainties $\sigma$ of the flux density $F_\lambda$ measured in a spectrograph are a combination of the non-photonic background term $\sigma_\mathrm{offset}$ and the photon noise that scales linearly with the square root of $F_\lambda$:

\begin{equation}\label{noise_eq}
\sigma^2 = \sigma_\mathrm{offset}^2 + \left(G \sqrt{F_\lambda}\right)^2 \mathrm{.}
\end{equation}

These parameters can vary broadly between different instruments. Based on the observations of GJ~436 reported by \citet{2017A&A...605L...7L}, we found that, for HST/STIS, $\sigma_\mathrm{offset} \approx 4.15 \times 10^{-16}$ erg s$^{-1}$ cm$^{-2}$ \AA$^{-1}$ and $G \approx 3.74 \times 10^{-8}$ (erg s$^{-1}$ cm$^{-2}$ \AA$^{-1}$)$^{1/2}$ for one exposure of 50 minutes using the grating G140M. With these numbers, we estimate that the mean uncertainty added by geocoronal photon noise in a 50-minute long exposure of HST/STIS amounts to $4.7 \times 10^{-15}$ erg s$^{-1}$ cm$^{-2}$.

The factor $G$ in Eq. \ref{noise_eq} depends on the exposure time and several instrumental properties, such as mirror size, and the efficiencies of the optics and the detector. To realistically compute the expected uncertainties of the observable Ly$\alpha$ spectrum measured with HST/STIS and other instruments, we consider that

\begin{equation}\label{G_factor}
  G \propto \frac{1}{A_\mathrm{eff}\sqrt{t_\mathrm{exp}}}\ \mathrm{,}
\end{equation}where $A_\mathrm{eff}$ is the effective area of the instrument (which is the end-to-end throughput multiplied by the geometric area of an unobstructed circular aperture with the size of the mirror) and $t_\mathrm{exp}$ is the exposure time.

The uncertainty $\sigma_S$ of the mean excess absorption depth $S$ is obtained by propagating the uncertainties of $F_{\mathrm{in}}$, $F_{\mathrm{out}}$ and $a_\mathrm{pl}$:

\begin{equation}
 \sigma_S^2 = \left(\frac{F_\mathrm{in}}{F_\mathrm{out}}\right)^2\left[\left(\frac{\sigma_\mathrm{in}}{F_\mathrm{in}}\right)^2 + \left(\frac{\sigma_\mathrm{out}}{F_\mathrm{out}}\right)^2\right] + \sigma_a^2 \mathrm{.}
\end{equation}If we assume that $F_\mathrm{in} \approx F_\mathrm{out}$ and ignoring the uncertainty $\sigma_a$ of $a_\mathrm{pl}$ for now, then $\sigma_S$ simplifies to

\begin{equation}\label{sigma_S_eq}
  \sigma_S = \frac{\sigma}{F} \sqrt{2} \mathrm{.}
\end{equation}

\subsubsection{HST/STIS}

According to the manual of HST/STIS\footnote{\footnotesize{Available at \url{http://www.stsci.edu/hst/stis/performance/throughput}.}}, the throughput of the instrument using the grating G140M (the usual to perform Ly$\alpha$ spectroscopy of M dwarfs) is $\sim$1.5\%, resulting in an effective area of 680 cm$^2$ around 1216 \AA.

\begin{figure*}
  \centering
  \includegraphics[width=1.1\hsize]{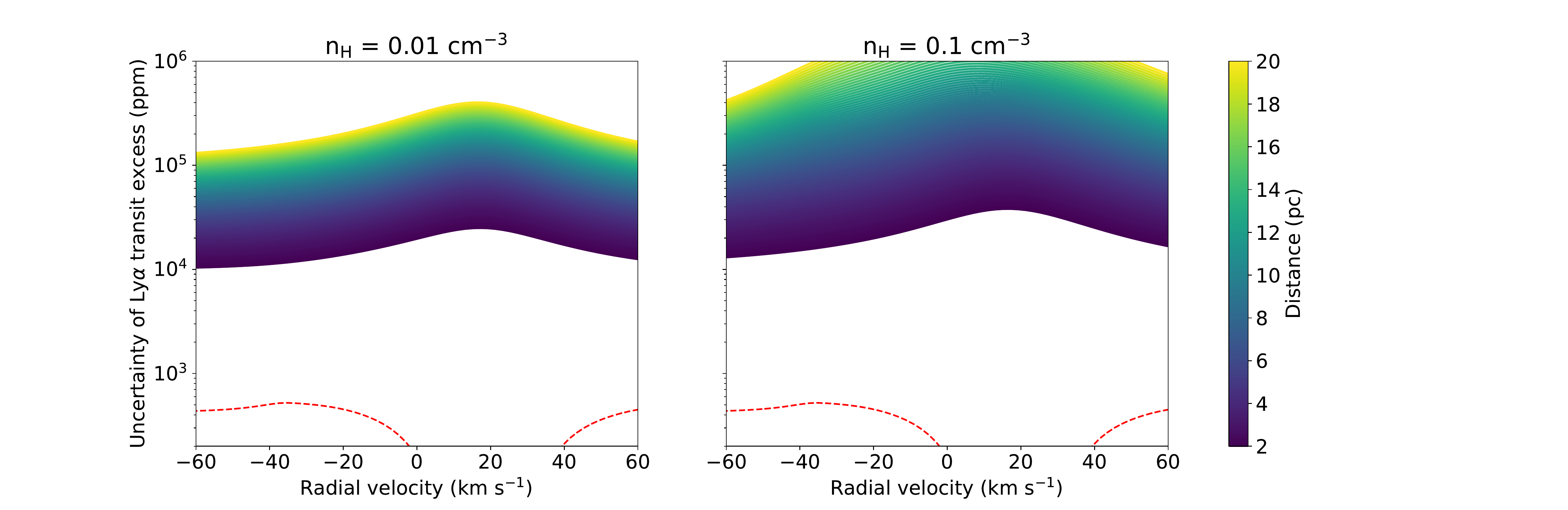}
  \caption{Expected uncertainty of the Lyman-$\alpha$ flux of M dwarfs in function of their distance in a single exposure of HST/STIS using the G140M grating and for two different densities of \ion{H}{I} in the ISM. The flux was computed between Doppler velocities [-50, 10] km s$^{-1}$ and we assumed the radial velocity of the ISM to be null in the telescope's rest frame. The dashed line represents the excess transit absorption level in a 0.1 R$_\odot$ M dwarf.}
        \label{unc_HST}
  \end{figure*}

In the regime where radiation pressure dominates the dynamics of an exosphere, we expect the velocities of the H particles in the upper atmosphere of the planet to be between the limits we described in Sect. \ref{dens} (usually between -40 and 0 km s$^{-1}$ in the rest frame of the star). Thus, depending on the difference between the radial velocity of the host star and that of the ISM cloud in the line of sight, the exospheric absorption signal may fall inside (totally or partially) or outside of the wavelength region where there is no flux to be measured (i.e. no information can be obtained).

Assuming no slit losses, we used Eqs. \ref{noise_eq}, \ref{G_factor} and \ref{sigma_S_eq} to estimate the uncertainty of the mean excess absorption in Ly$\alpha$ flux of M dwarfs between -50 and 10 km s$^{-1}$ in the stellar rest frame (the range where we expect the exospheric absorption signal to be located). The estimates were performed for a range of stellar stellar radial velocities and distances, assuming the Ly$\alpha$ profile to be similar to GJ~436, and we plot the results in Fig. \ref{unc_HST}.

For a small M dwarf with $R = 0.1$ R$_\odot$, the expected excess transit depth between Doppler velocities [-50, 10] km s$^{-1}$ is approximately 400 ppm for HST/STIS (dashed line in Fig. \ref{unc_HST}), and such a precision is beyond the capabilities of the instrument. The increase in flux uncertainty between radial velocities 0 and 40 km s$^{-1}$ is due to the ISM absorption. Furthermore, we note that, in the case of a large density of \ion{H}{I} in the line of sight (as seen in the right panel of Fig. \ref{unc_HST}), little to no information about the flux can be measured for M dwarfs beyond $\sim$16 pc and radial velocities between -20 and 30 km s$^{-1}$ if we use HST/STIS. These results are compatible with the precision obtained for the Ly$\alpha$ flux of TRAPPIST-1 using HST/STIS data \citep{2017AJ....154..121B}.

\subsubsection{LUVOIR/LUMOS and LUVOIR/POLLUX}

The detection of an exosphere around an Earth-like planet may require the use of a larger telescope and improved efficiencies to decrease the uncertainties of the Ly$\alpha$ flux. For comparison, the expected effective areas of LUVOIR using the spectrograps POLLUX and LUMOS are, respectively, 2~000 (1~100) and 130~000 (73~100) cm$^2$, assuming the design with a diameter of 12 (9) m (K. France, private communication; see a summary of instrumental properties in Table \ref{instr_prop}).

\begin{table*}
  \caption{Properties of the instruments.}
  \label{instr_prop}
  \centering
  \begin{tabular}{l c c c}
  \hline\hline
  \multirow{2}{*}{Properties near Ly$\alpha$} & HST/STIS & LUVOIR-A(B)/LUMOS & LUVOIR-A(B)/POLLUX \\
   & G140M & G120M & \\
  \hline
  Resolving power & $10\ 000$ & $44\ 000$ & $120\ 000$ \\
  Effective area (cm$^2$) & $680$ & $130\ 000$ ($73\ 100$) & $2\ 000$ ($1\ 100$) \\
  Background rate (erg cm$^{-2}$ s$^{-1}$ \AA$^{-1}$) & $7 \times 10^{-17}$ & $2 \times 10^{-20}$ & ... \\
  Aperture size (arcsec) & $52 \times 0.1$ [slit] & $0.136 \times 0.068$ [shutter] & 0.03 [pinhole] \\
  \hline
  \end{tabular}
  \end{table*}

\begin{figure*}
  \centering
  \includegraphics[width=1.1\hsize]{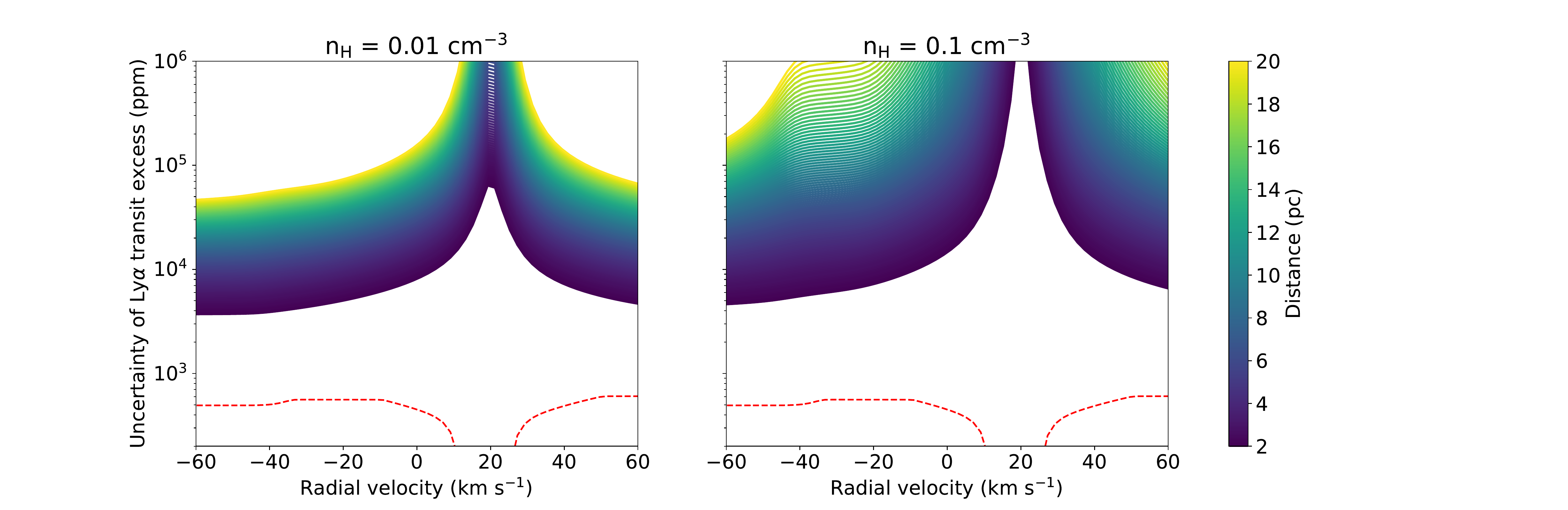} 
  \includegraphics[width=1.1\hsize]{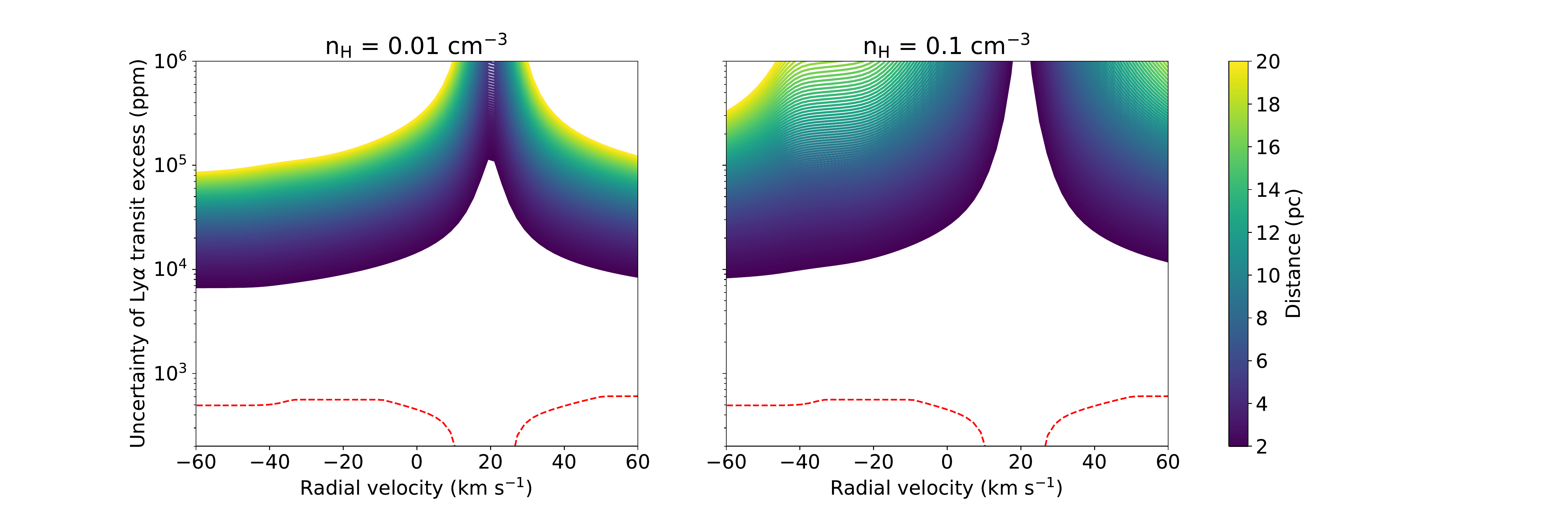} 
  \caption{Same as Fig. \ref{unc_HST}, but for LUVOIR/POLLUX. The top (lower) row corresponds to LUVOIR-A (-B).}
        \label{unc_LUVOIR_POLLUX}
  \end{figure*}

\begin{figure*}
    \centering
    \includegraphics[width=1.1\hsize]{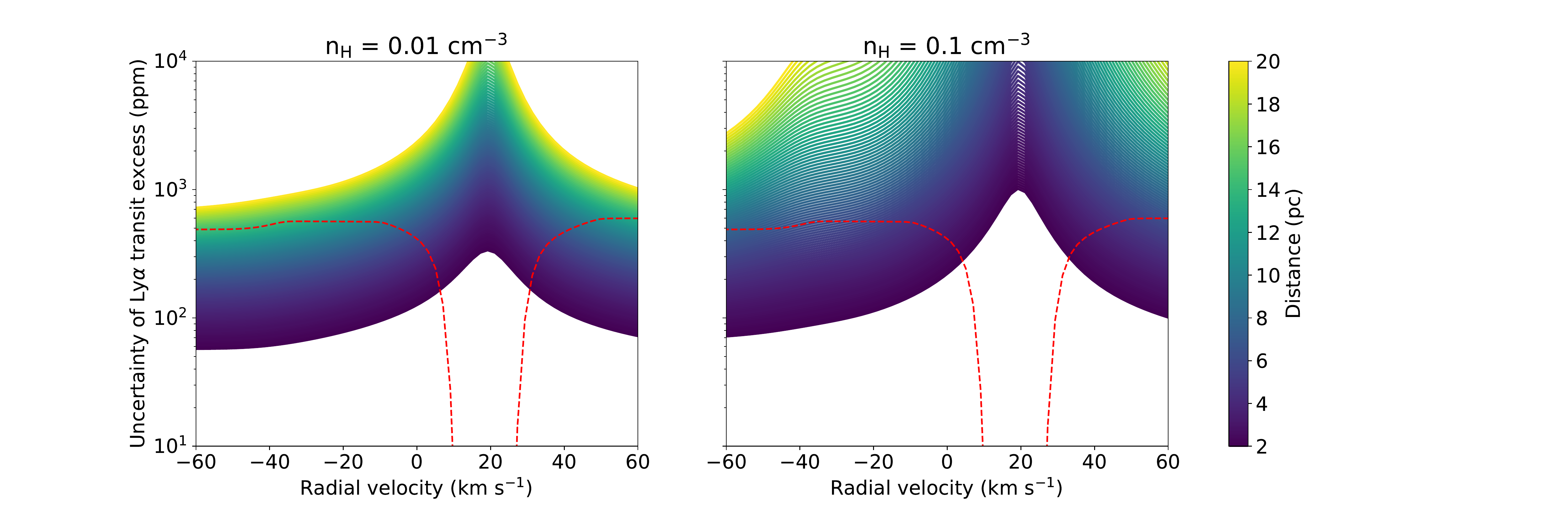} 
    \includegraphics[width=1.1\hsize]{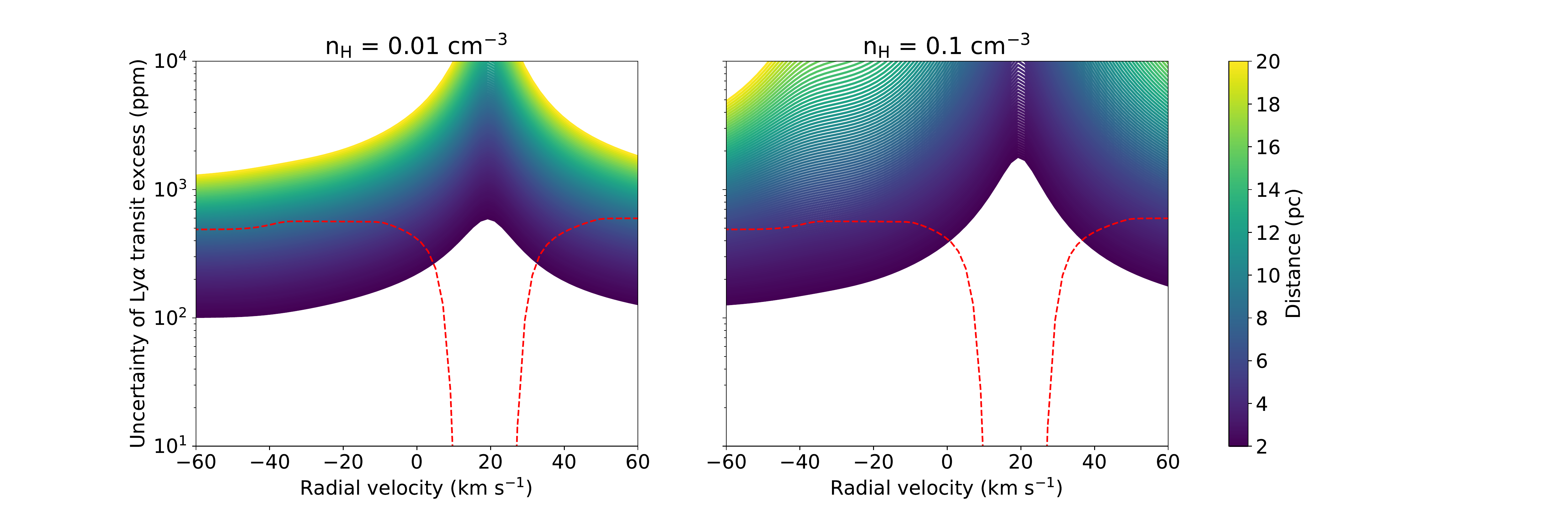} 
    \caption{Same as Fig. \ref{unc_HST}, but for LUVOIR/LUMOS and with a different scale in the y-axis.}
        \label{unc_LUVOIR_LUMOS}
    \end{figure*}

We performed the same tests with LUVOIR/LUMOS and LUVOIR/POLLUX\footnote{\footnotesize{For POLLUX we assumed the same instrumental background as LUMOS.}} to estimate the uncertainties they will yield when observing M dwarfs at different distances; the results are summarized in Figs. \ref{unc_LUVOIR_POLLUX} and \ref{unc_LUVOIR_LUMOS}. It is important to note that LUVOIR will reside at the Earth-Sun L2 point, thus avoiding contamination by geocoronal emission. It will, however, still contain astrophysical background contamination; based on the value $\sim$710 R reported by \citet{2017GeoRL..4411706K}, we estimate that the astrophysical background in Ly$\alpha$ to be $\sim$$2 \times 10^{-14}$ erg s$^{-1}$ cm$^{-2}$ arcsec$^{-2}$.

Our estimates show that, in a 50-minutes\footnote{\footnotesize{We chose 50 minutes so that the results are comparable to HST/STIS. The duration of an HZ exoplanet transit around an M dwarf is in this order of magnitude.}} exposure with LUVOIR, the uncertainties of the Ly$\alpha$ flux of M dwarfs up to a distance of 20 pc and up to radius 0.6 R$_\odot$ are always limited by the photon noise of the source, and well above the limit set by the photon noise of background astrophysical sources. The LUMOS spectrograph will have uncertainties two orders of magnitude lower than POLLUX owing to its larger effective area. POLLUX, on the other hand, provides an improvement of one order of magnitude in the flux precision when compared to HST/STIS.
  
By comparing the estimated uncertainty levels with the expected signal in Fig. \ref{t_depth}, one can conclude that LUVOIR/LUMOS will be better suited to perform the detection of an Earth-like exosphere transiting an M dwarf in Ly$\alpha$. Such a detection will, nonetheless, be limited to a certain distance, which will vary depending on the time we are willing to invest. If we consider that 30 transits is a reasonable time investment for a single target, then this distance limit should be 15-20 pc in the most favorable scenarios.

\subsection{Analysis of a synthetic dataset with injected noise}\label{synth_cases}

\begin{figure*}
  \centering
  \begin{tabular}{cc}
  \includegraphics[width=0.48\hsize]{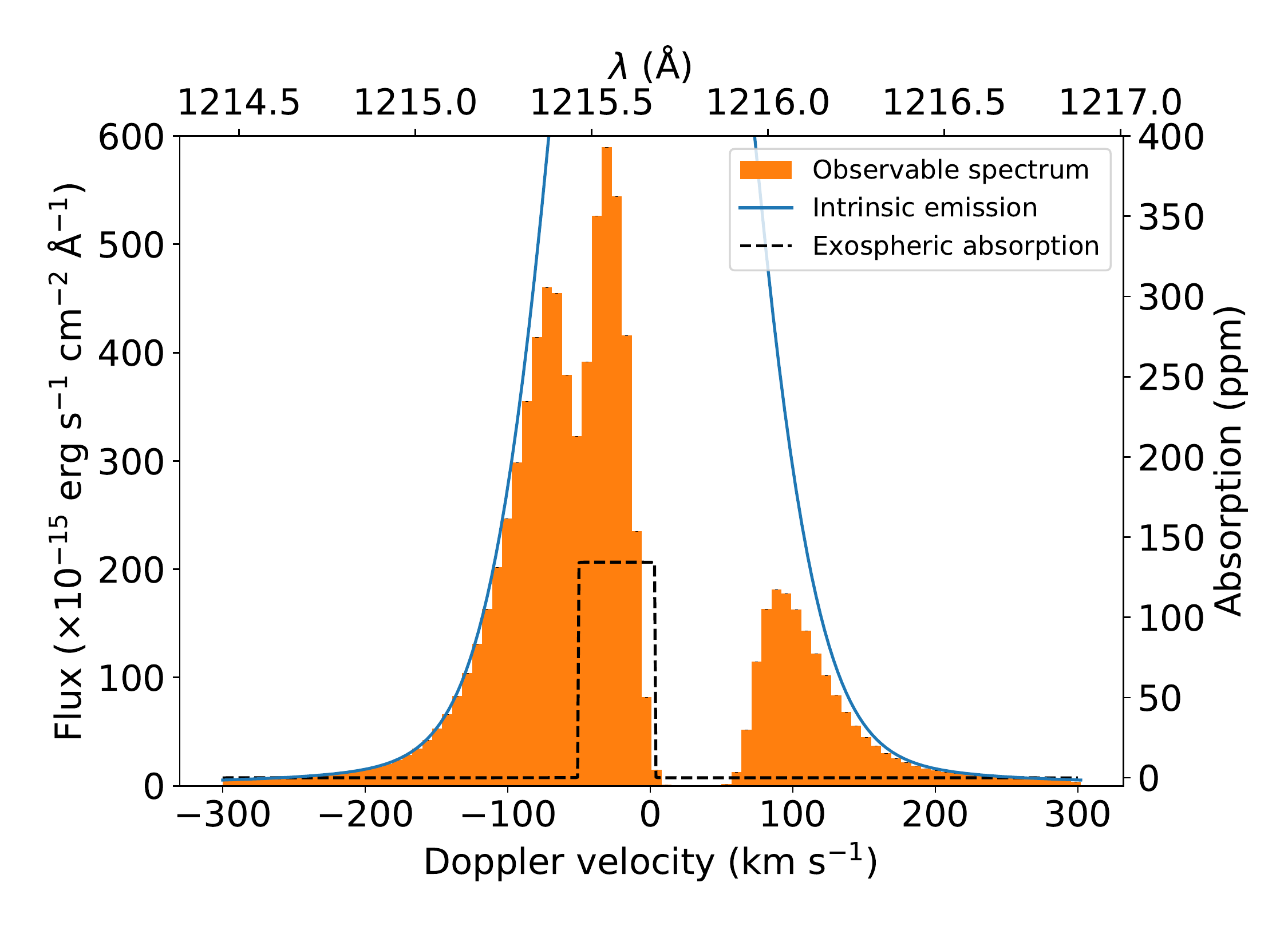} & \includegraphics[width=0.48\hsize]{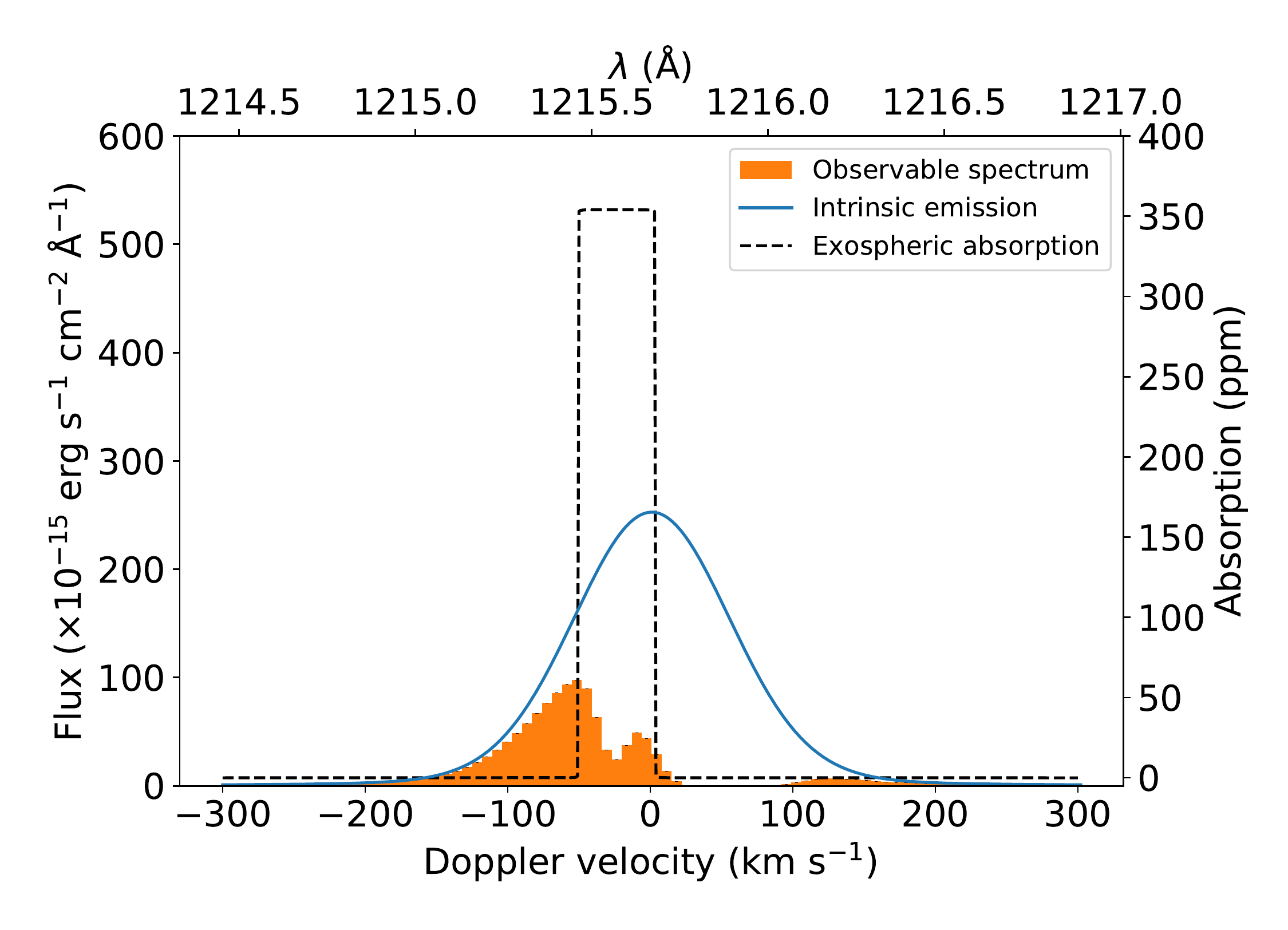} \\
  \end{tabular}
  \caption{Synthetic spectra used in the test case 1 (left) and 2 (right). The scales in the y-axes are kept the same to highlight the difference between the observable fluxes.}
        \label{star_spectrum}
  \end{figure*}

To illustrate the point of this manuscript, we analyze two test-cases with synthetic datasets and a well-established analysis strategy used in previous studies of FUV transmission spectroscopy. In case 1, we produce the Ly$\alpha$ spectrum of a M dwarf with $R = 0.2$ R$_\odot$, $V_\mathrm{star} = -30$~km~s$^{-1}$, $d = 5$ pc, and simulate the observable spectra with realistic injected noise based on the instrumental design of LUVOIR/LUMOS. In case 2, to simulate a case similar to TRAPPIST-1, we change to stellar radius to 0.12 R$_\odot$, distance to $d = 12$ pc and $V_\mathrm{star} = -55$~km~s$^{-1}$. The Ly$\alpha$ synthetic spectra of the test-cases are shown in Fig. \ref{star_spectrum}.

\begin{figure*}
  \centering
  \begin{tabular}{cc}
  \includegraphics[width=0.48\hsize]{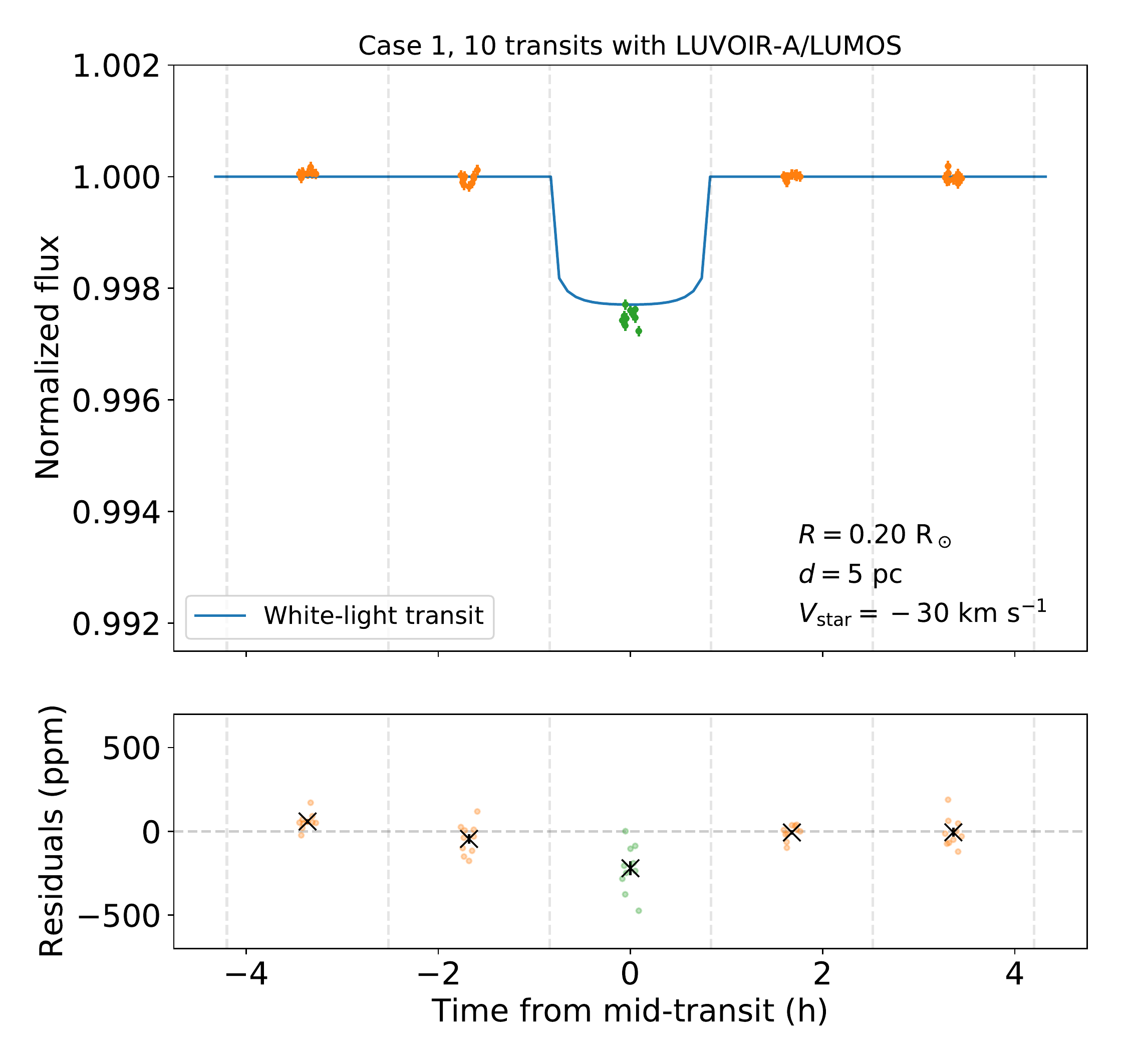} & \includegraphics[width=0.48\hsize]{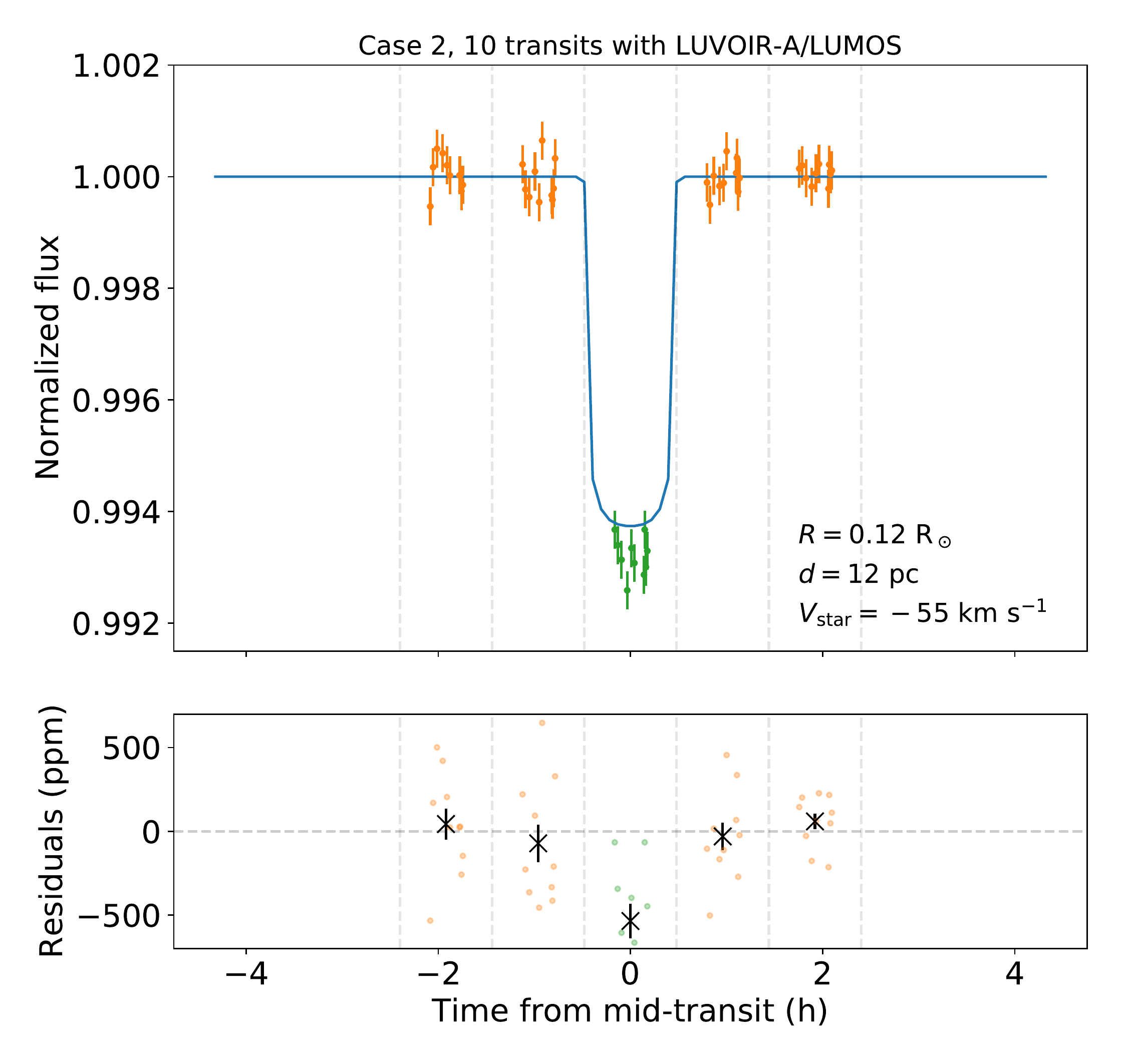} \\
  \includegraphics[width=0.48\hsize]{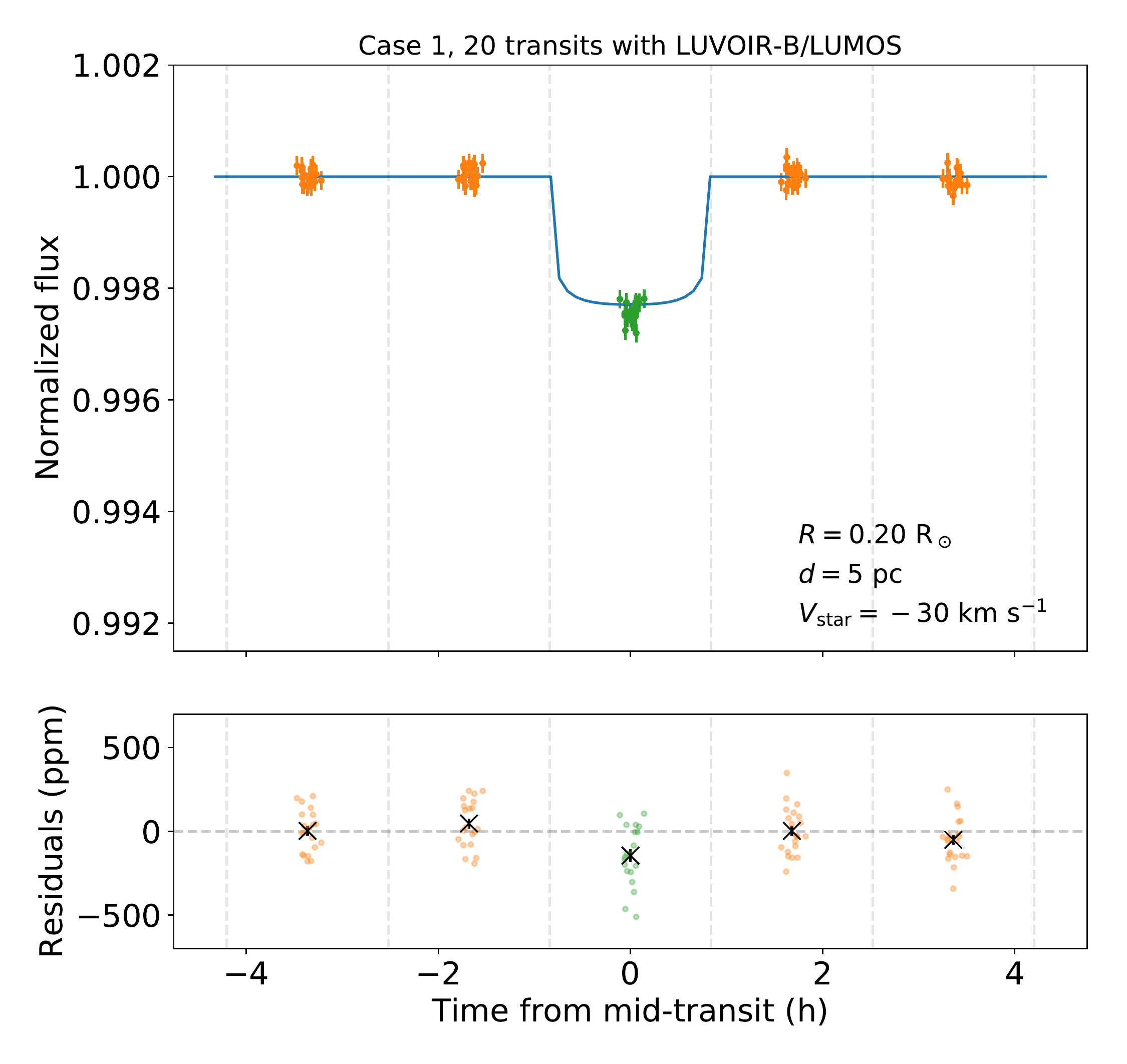} & \includegraphics[width=0.48\hsize]{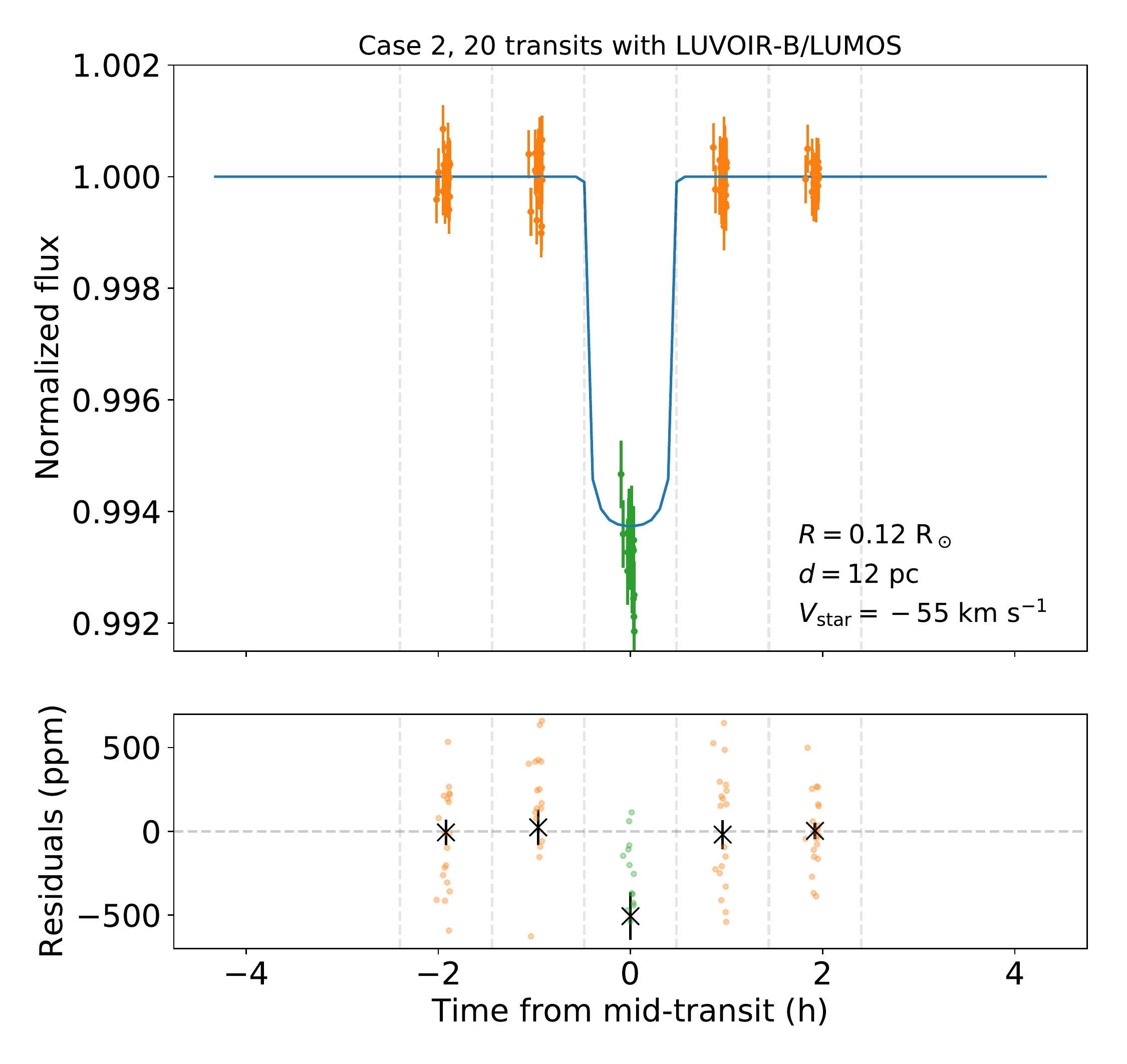} \\
  \end{tabular}
  \caption{Transit light curves of our test cases. The vertical dashed lines separate the different exposures (5 in total for each transit; the exposure time for each data point is equal to the transit duration).}
        \label{lc}
  \end{figure*}

We produce light curves of the Ly$\alpha$ flux of the star by integrating its flux density between the ranges [-50, 10] km s$^{-1}$ in the stellar reference frame, assuming that all transit exposures happened at roughly the same phases of the planetary orbit. The exposure time is set by the duration of the optical transit, which should be around 100 (60) minutes in the case of an Earth-sized planet orbiting in the inner edge of the HZ of a $R = 0.20$ (0.12) R$_\odot$ M dwarf. The resulting light curves are shown in Fig. \ref{lc}.

The white-light absorption dominates the total Ly$\alpha$ occultation for an Earth-like exoplanet transiting a M dwarf, as expected. However, with the accumulation of several transit light curves with LUVOIR/LUMOS, there is a visible excess absorption in the mid-transit exposures ($\sim$200 ppm for case 1 and 500 ppm for case 2, values that are comparable to the expected transit depth excess seen in Fig. \ref{t_depth}), assuming that the white-light radius of the transiting exoplanet is known perfectly. For a reference, the precision of TESS light curves for bright targets ($T_\mathrm{mag} < 8$) is on the order of 100 ppm \citep{2015ApJ...809...77S, 2018arXiv180302316O}, which is similar to the signal we are trying to detect; thus, an extensive photometric follow-up on Earth-sized candidates from \emph{TESS} may be necessary to obtain a precise white-light radius before attempting a Ly$\alpha$ transmission spectrum; the satellite \emph{CHEOPS} \citep{2013EPJWC..4703005B} will be a crucial instrument to obtain precise radii of exoplanets.

The Ly$\alpha$ transit excess absorption obtained in the noise-injected test cases are shown in Table \ref{transit_excess}. In general, we show that it is possible to perform the detections within a resonable time investment (10-20 transits) using LUVOIR/LUMOS and obtain a significance of 5$\sigma$ for a targets at 5 pc or 12 pc.

  \begin{table}
    \caption{Mid-transit excess absorption obtained for cases 1 and 2 using noise-injected data based on the specifications of LUVOIR/LUMOS.}
    \label{transit_excess}
    \centering
    \begin{tabular}{l c c c}
    \hline\hline
    \multirow{2}{*}{} & \multicolumn{2}{c}{Transit excess (ppm)} & \multirow{2}{*}{\# of transits} \\
    & Case 1 & Case 2 & \\
    \hline
    LUVOIR-A & $220 \pm 42$ & $535 \pm 103$ & 10 \\
    LUVOIR-B & $145 \pm 39$ & $507 \pm 142$ & 20 \\
    \hline
    \end{tabular}
    \end{table}

\subsection{Comparison with previous studies}

A recent report by \citet{2018ExA...tmp....8C} showed that detecting exospheric hydrogen in terrestrial planets is feasible for nearby M dwarf targets, provided that the telescope is large enough (4-8 primary mirror). Our results corroborate this conclusion of \citeauthor{2018ExA...tmp....8C}, but we found that the detectability depends not only on the telescope size, but it is also highly-dependent on the intrinsic properties of the host star, such as radius and the radial velocity difference of the system. In fact, stars with favorable properties may allow us to search for H-rich exospheres in a range of distances up to 20 pc (see Sect. \ref{toi}).

There are other, more direct but not necessarily more efficient, ways to search for evidence of atmospheric water in Earth-sized planets. Transmission spectroscopy using \emph{JWST} \citep{2016ApJ...817...17G} offers the most obvious choice. In particular, \citet{2016MNRAS.461L..92B} showed that it is possible to detect water in the atmospheres of some planets in the TRAPPIST-1 system using JWST, but it would require 60 transits for a clear detection. Another strategy that may become available in the future is high-dispersion spectroscopy coupled with high-resolution imaging \citep[e.g.,][]{2015A&A...576A..59S, 2017A&A...599A..16L}, but such techniques may be limited to very nearby systems (up to only a few pc, such as Proxima b). In addition, high-dispersion spectroscopy with the extremely large telescopes may provide a window to search for biosignatures, such as oxygen, in rocky nearby planets up to several pc in distance \citep{2013ApJ...764..182S}.

\section{Targets of interest}\label{toi}

Our results from Sects. \ref{lya-m-dwarfs} and \ref{obs_strategy} show that the best targets to search for the signal of exospheric absorption in an Earth-like exoplanet are small M dwarfs within $\sim$15 pc. These limits can be stretched or shrunk depending on the radial velocity of the star and the \ion{H}{I} density of the ISM in the line of sight. To date, the nearest M dwarfs with known Earth-sized transiting planets are LHS 1140 \citep{2017Natur.544..333D, 2018arXiv180800485M} and TRAPPIST-1 \citep{2016Natur.533..221G, 2017Natur.542..456G}. We were able to obtain a detection in simulated data for a TRAPPIST-1-like system, but the distance and radial velocity of LHS 1140 (15 pc and -13.2 km s$^{-1}$, respectively) may render the latter a difficult target for the search of an Earth-like exosphere.

In the case of nearby solar-type stars, such as $\alpha$ Cen AB and 18 Sco, the most significant barrier to perform this search is that the inner edge of their HZ sits too far from the host star, rendering orbital periods in the order of several months \citep{1993Icar..101..108K} and severely decreasing the transit probability of a potential exoplanet.

More Earth-sized planets amenable to exospheric characterization are likely going to be discovered by \emph{TESS} \citep{2014SPIE.9143E..20R}. The mission will detect several new transiting exoplanets in the solar neighborhood, particularly around M dwarfs in a brighter regime than the \emph{Kepler} and \emph{K2} missions. Based on the Catalogue of Simulated \emph{TESS} Detections \citep{2015ApJ...809...77S}, we estimate that planets amenable to atmospheric escape detection -- namely those around stars within 50 pc -- will amount to a sample of $\sim 130$ planets. Of these, $\sim 30$ will have radii between 0.7 and 1.5 R$_\oplus$, of which $\sim 10$ will have a similar level of irradiation as the Earth does. A final cut to a distance of 15 pc results that only $\sim 4$ of these predicted transiting planets will be similar to Earth in size and irradiation level, and viably be amenable to exospheric characterization using LUVOIR.

In this manuscript we limited ourselves to consider only quiet M dwarfs and exoplanets with an exosphere similar to the Earth's, which do not provide the most favorable conditions to produce evidence for water in the lower atmosphere of a rocky exoplanet. Other targets that can potentially be of interest are active M dwarfs that host transiting exoplanets, which are known to display strong Ly$\alpha$ emission and thus would likely produce stronger exospheric absorption signal.

\begin{figure}
  \centering
  \includegraphics[width=\hsize]{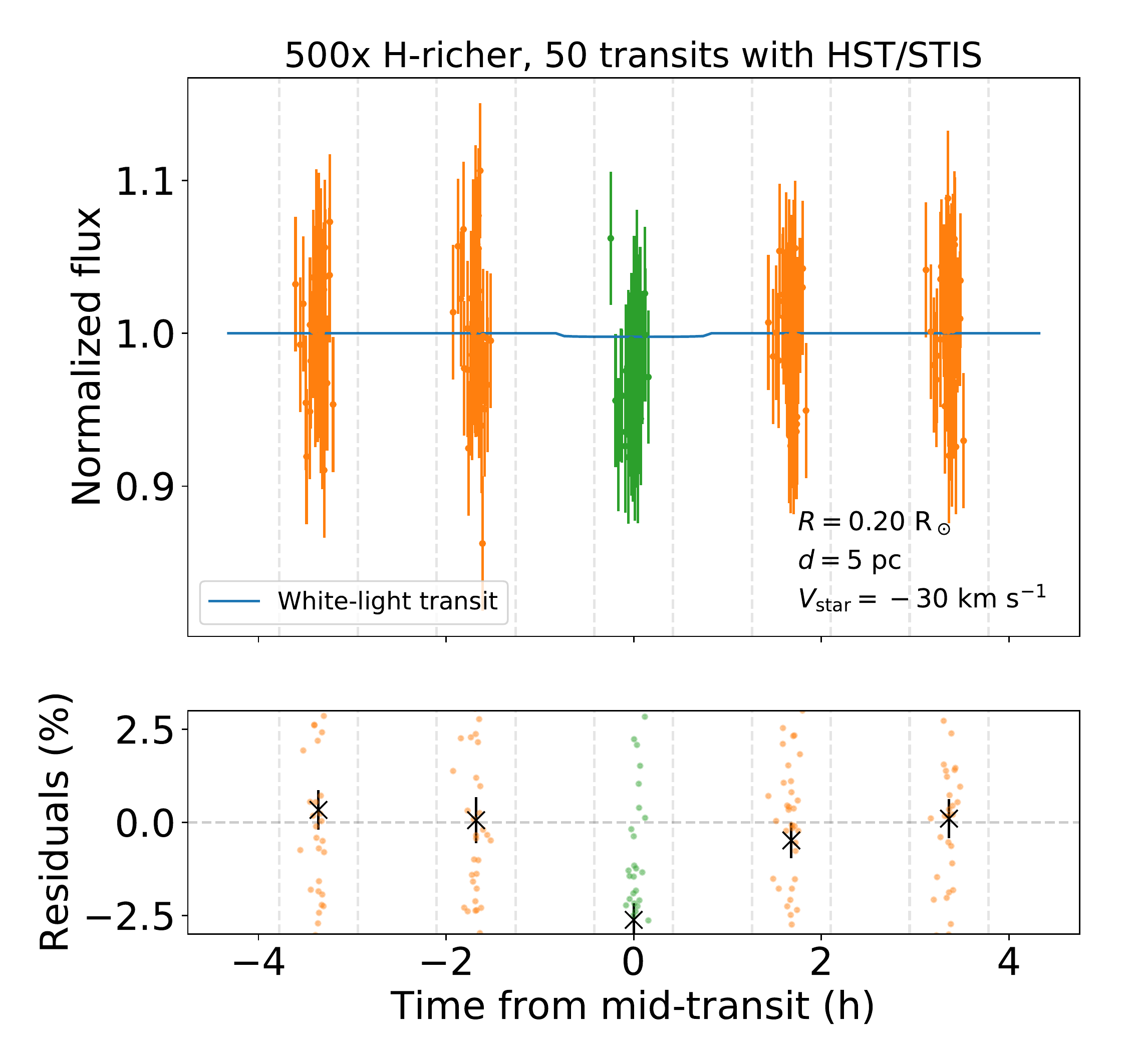}
  \caption{Transit light curves of test case 1 with an exosphere 500 times denser than the Earth's. The Ly$\alpha$ transit excess is detected reliably within 50 transits with HST/STIS.}
        \label{lc_hst}
  \end{figure}

Exoplanets with denser exospheres will produce a stronger absorption signal in Ly$\alpha$, and our simulations indicate that the signal is proportional to the overall density of the exosphere, as long as it remains optically thin. Although simulating atmospheric escape in water-rich planets is not in the scope of our study, we expect that such planets could have exospheres orders of magnitude more dense than the Earth, and could be potentially detectable with HST/STIS. We computed the synthetic spectrum of the star from case 1 (see Sect. \ref{synth_cases}) as if observed with HST/STIS, and produced an exosphere model with densities 500 times larger than the Earth's to test the limit we can observe with this instrument: our simulation indicates that a signal can be detected with significance 4$\sigma$ during 50 transits with HST/STIS, assuming that the exosphere has the same shape as the Earth's (see Fig. \ref{lc_hst}).

Theoretical models of planet formation predict a pathway to form deep ocean planets, in which a significant part of their mass (up to $\sim$50\%) is composed of water \citep[e.g.,][]{2004Icar..169..499L}. Furthermore, based on demographics of exoplanets observed with Kepler, it is also possible that many small planets are dense enough to be rocky, but may harbour thick envelopes rich in volatiles \citep{2015ApJ...801...41R}. Such water- and volatile-rich planets are prime targets for the search of H-rich exospheres using the UV capabilities of HST.

\section{Conclusions}\label{conclusions}

Studying the upper atmospheres of extrasolar planets will be an important tool to move forward towards understanding habitability and planetary evolution. The region of the upper atmosphere known as exosphere is mainly composed of neutral hydrogen that escapes from the planet driven by the high-energy irradiation from the host star and, for an Earth-like planet, it can be a product of the presence of water in the lower atmosphere. In extrasolar planets, this feature has only been confirmed for giant planets thus far, but the Earth is known to possess its own exosphere. In this work, we sought to investigate if it is possible to observe the exosphere of an Earth-like planet orbiting nearby M dwarfs using transit spectroscopy with the Ly$\alpha$ line. To perform such tests, we employ realistic constraints in the instrumental limitations of current and future space missions, namely the HST and LUVOIR, and the empirical model of the geocorona proposed by \citet{2017GeoRL..4411706K}.

The observable Ly$\alpha$ flux of stars scales most strongly with their distance, owing to ISM absorption in the line of sight. We found that the intrinsic Ly$\alpha$ flux attenuation $\delta$ with distance $d$ follows an exponential relation, and will also depend on the density of \ion{H}{I} atoms in the line of sight. This relation is useful not only to achieve our particular objectives, but also to estimate the observed Ly$\alpha$ flux in any star in the solar neighborhood. Further, we obtained several relations between the excess absorption caused by a transiting exosphere during the mid-transit of an Earth-like planet. These relations are complex, owing to the fact that they depend on the difference between the radial velocity of the star and that of the ISM in the line of sight, and also on the instrumental setup.

We presented an extensive and detailed discussion on the observational strategies necessary to perform the above-mentioned detection in Section \ref{obs_strategy}, which we summarize below:

\begin{itemize}
\item The space telescope LUVOIR will be able to characterize exospheres in Earth-like planets using the LUMOS spectrograph. The number of transits necessary to obtain a reliable detection varies strongly depending on the target, but we can set a flexible limit at distances up to 15 pc for 30 transits. The astrophysical background contamination will not be a limitation for nearby targets, neither will the Earth's own exosphere.
\item We analyzed two test cases with noise-injected, simulated data: case 1 had $R = 0.20$ R$_\odot$, $d = 5$ pc and $\Delta V = -30$ km s$^{-1}$; for case 2, we used a system similar to TRAPPIST-1, in which $R = 0.12$ R$_\odot$, $d = 10$ pc and $\Delta V = -55$ km s$^{-1}$. In both cases, even though they did not involve the most favorable conditions, we obtained reliable detections within 10 (20) transits with LUVOIR-A (-B).
\item Although an Earth-like exosphere is not observable in Ly$\alpha$ spectroscopy with HST/STIS, we were able to obtain a reliable detection using noise-injected data in the case of an exosphere 500 times more dense in 50 transits with this instrument, for a target at a distance of 5 pc.
\end{itemize}

The nearest M dwarf with transiting planets is TRAPPIST-1, and its distance of 12 pc renders it a clear target for Ly$\alpha$ transmission spectroscopy with LUVOIR/LUMOS. LHS 1140, another nearby M dwarf that hosts transiting planets is a more challenging target owing to its unfavorable radial velocity. We did not consider active M dwarfs in this study but, owing to their stronger Ly$\alpha$ emission than inactive ones, they likely have more extended limits in distance than inactive M dwarfs. The \emph{TESS} mission will expand the search for transiting planets around nearby M dwarfs to the whole sky, which will likely yield more viable targets for atmospheric escape characterization. Future studies in this direction will benefit from performing simulations of neutral hydrogen exospheres \citep[e.g., using the EVE models;][]{2013A&A...557A.124B, 2016A&A...591A.121B} around rocky planets with denser atmospheres (such as ocean or steam worlds) orbiting M dwarfs; these stars have a more intense high-energy output than Sun-like stars and likely produce denser exospheres and shape them into comet-like tails by radiation pressure, increasing the detectability.

\begin{acknowledgements}
  This project has received funding from the European Research Council (ERC) under the European Union’s Horizon 2020 research and innovation programme (project {\sc Four Aces}; grant agreement No 724427), and it has been carried out in the frame of the National Centre for Competence in Research PlanetS supported by the Swiss National Science Foundation (SNSF). The authors thank B. Wood for providing us the intrinsic Ly$\alpha$ profile of Proxima Centauri that was used in a previous version of this manuscript, and acknowledge the information about the LUMOS and POLLUX spectrographs kindly provided by K. France and C. Neiner, respectively. We also thank A. Su\'arez Mascare\~no, J. Seidel, C. Lovis and the anonymous referee for providing valuable input that helped improve the manuscript. This research made use of the software SciPy \citep{scipy_ref}, Astropy \citep{2013A&A...558A..33A}, Matplotlib \citep{Hunter:2007}, Jupyter \citep{Kluyver:2016aa}, \texttt{emcee} \citep{2013PASP..125..306F} and \texttt{batman} \citep{2015PASP..127.1161K}, as well as the online databases SIMBAD and VizieR \citep{2000A&AS..143....9W, 2000A&AS..143...23O}, operated at CDS, Strasbourg, France.
\end{acknowledgements}

\bibliographystyle{aa}
\bibliography{biblio.bib}

\begin{thebibliography}{71}
\expandafter\ifx\csname natexlab\endcsname\relax\def\natexlab#1{#1}\fi

\bibitem[{{Astropy Collaboration} {et~al.}(2013){Astropy Collaboration},
  {Robitaille}, {Tollerud}, {Greenfield}, {Droettboom}, {Bray}, {Aldcroft},
  {Davis}, {Ginsburg}, {Price-Whelan}, {Kerzendorf}, {Conley}, {Crighton},
  {Barbary}, {Muna}, {Ferguson}, {Grollier}, {Parikh}, {Nair}, {Unther},
  {Deil}, {Woillez}, {Conseil}, {Kramer}, {Turner}, {Singer}, {Fox}, {Weaver},
  {Zabalza}, {Edwards}, {Azalee Bostroem}, {Burke}, {Casey}, {Crawford},
  {Dencheva}, {Ely}, {Jenness}, {Labrie}, {Lim}, {Pierfederici}, {Pontzen},
  {Ptak}, {Refsdal}, {Servillat}, \& {Streicher}}]{2013A&A...558A..33A}
{Astropy Collaboration}, {Robitaille}, T.~P., {Tollerud}, E.~J., {et~al.} 2013,
  \aap, 558, A33

\bibitem[{{Barstow} \& {Irwin}(2016)}]{2016MNRAS.461L..92B}
{Barstow}, J.~K. \& {Irwin}, P.~G.~J. 2016, \mnras, 461, L92

\bibitem[{{Bolcar} {et~al.}(2017){Bolcar}, {Aloezos}, {Bly}, {Collins},
  {Crooke}, {Dressing}, {Fantano}, {Feinberg}, {France}, {Gochar}, {Gong},
  {Hylan}, {Jones}, {Linares}, {Postman}, {Pueyo}, {Roberge}, {Sacks},
  {Tompkins}, \& {West}}]{2017SPIE10398E..09B}
{Bolcar}, M.~R., {Aloezos}, S., {Bly}, V.~T., {et~al.} 2017, in Society of
  Photo-Optical Instrumentation Engineers (SPIE) Conference Series, Vol. 10398,
  Society of Photo-Optical Instrumentation Engineers (SPIE) Conference Series,
  1039809

\bibitem[{{Bolmont} {et~al.}(2017){Bolmont}, {Selsis}, {Owen}, {Ribas},
  {Raymond}, {Leconte}, \& {Gillon}}]{2017MNRAS.464.3728B}
{Bolmont}, E., {Selsis}, F., {Owen}, J.~E., {et~al.} 2017, \mnras, 464, 3728

\bibitem[{{Bourrier} {et~al.}(2017{\natexlab{a}}){Bourrier}, {de Wit},
  {Bolmont}, {Stamenkovi{\'c}}, {Wheatley}, {Burgasser}, {Delrez}, {Demory},
  {Ehrenreich}, {Gillon}, {Jehin}, {Leconte}, {Lederer}, {Lewis}, {Triaud}, \&
  {Van Grootel}}]{2017AJ....154..121B}
{Bourrier}, V., {de Wit}, J., {Bolmont}, E., {et~al.} 2017{\natexlab{a}}, \aj,
  154, 121

\bibitem[{{Bourrier} {et~al.}(2017{\natexlab{b}}){Bourrier}, {Ehrenreich},
  {Allart}, {Wyttenbach}, {Semaan}, {Astudillo-Defru}, {Gracia-Bern{\'a}},
  {Lovis}, {Pepe}, {Thomas}, \& {Udry}}]{2017A&A...602A.106B}
{Bourrier}, V., {Ehrenreich}, D., {Allart}, R., {et~al.} 2017{\natexlab{b}},
  \aap, 602, A106

\bibitem[{{Bourrier} {et~al.}(2017{\natexlab{c}}){Bourrier}, {Ehrenreich},
  {King}, {Lecavelier des Etangs}, {Wheatley}, {Vidal-Madjar}, {Pepe}, \&
  {Udry}}]{2017A&A...597A..26B}
{Bourrier}, V., {Ehrenreich}, D., {King}, G., {et~al.} 2017{\natexlab{c}},
  \aap, 597, A26

\bibitem[{{Bourrier} {et~al.}(2015){Bourrier}, {Ehrenreich}, \& {Lecavelier des
  Etangs}}]{2015A&A...582A..65B}
{Bourrier}, V., {Ehrenreich}, D., \& {Lecavelier des Etangs}, A. 2015, \aap,
  582, A65

\bibitem[{{Bourrier} {et~al.}(2018){Bourrier}, {Ehrenreich}, {Lecavelier des
  Etangs}, {Louden}, {Wheatley}, {Wyttenbach}, {Vidal-Madjar}, {Lavie}, {Pepe},
  \& {Udry}}]{2018A&A...615A.117B}
{Bourrier}, V., {Ehrenreich}, D., {Lecavelier des Etangs}, A., {et~al.} 2018,
  \aap, 615, A117

\bibitem[{{Bourrier} {et~al.}(2017{\natexlab{d}}){Bourrier}, {Ehrenreich},
  {Wheatley}, {Bolmont}, {Gillon}, {de Wit}, {Burgasser}, {Jehin}, {Queloz}, \&
  {Triaud}}]{2017A&A...599L...3B}
{Bourrier}, V., {Ehrenreich}, D., {Wheatley}, P.~J., {et~al.}
  2017{\natexlab{d}}, \aap, 599, L3

\bibitem[{{Bourrier} \& {Lecavelier des Etangs}(2013)}]{2013A&A...557A.124B}
{Bourrier}, V. \& {Lecavelier des Etangs}, A. 2013, \aap, 557, A124

\bibitem[{{Bourrier} {et~al.}(2016){Bourrier}, {Lecavelier des Etangs},
  {Ehrenreich}, {Tanaka}, \& {Vidotto}}]{2016A&A...591A.121B}
{Bourrier}, V., {Lecavelier des Etangs}, A., {Ehrenreich}, D., {Tanaka}, Y.~A.,
  \& {Vidotto}, A.~A. 2016, \aap, 591, A121

\bibitem[{{Broeg} {et~al.}(2013){Broeg}, {Fortier}, {Ehrenreich}, {Alibert},
  {Baumjohann}, {Benz}, {Deleuil}, {Gillon}, {Ivanov}, {Liseau}, {Meyer},
  {Oloffson}, {Pagano}, {Piotto}, {Pollacco}, {Queloz}, {Ragazzoni}, {Renotte},
  {Steller}, \& {Thomas}}]{2013EPJWC..4703005B}
{Broeg}, C., {Fortier}, A., {Ehrenreich}, D., {et~al.} 2013, in European
  Physical Journal Web of Conferences, Vol.~47, European Physical Journal Web
  of Conferences, 03005

\bibitem[{{Carruthers} \& {Page}(1972)}]{1972Sci...177..788C}
{Carruthers}, G.~R. \& {Page}, T. 1972, Science, 177, 788

\bibitem[{{Chamberlain}(1963)}]{1963P&SS...11..901C}
{Chamberlain}, J.~W. 1963, \planss, 11, 901

\bibitem[{{Deming} \& {Seager}(2017)}]{2017JGRE..122...53D}
{Deming}, L.~D. \& {Seager}, S. 2017, Journal of Geophysical Research
  (Planets), 122, 53

\bibitem[{{Dittmann} {et~al.}(2017){Dittmann}, {Irwin}, {Charbonneau},
  {Bonfils}, {Astudillo-Defru}, {Haywood}, {Berta-Thompson}, {Newton},
  {Rodriguez}, {Winters}, {Tan}, {Almenara}, {Bouchy}, {Delfosse}, {Forveille},
  {Lovis}, {Murgas}, {Pepe}, {Santos}, {Udry}, {W{\"u}nsche}, {Esquerdo},
  {Latham}, \& {Dressing}}]{2017Natur.544..333D}
{Dittmann}, J.~A., {Irwin}, J.~M., {Charbonneau}, D., {et~al.} 2017, \nat, 544,
  333

\bibitem[{{Ehrenreich} {et~al.}(2015){Ehrenreich}, {Bourrier}, {Wheatley},
  {Lecavelier des Etangs}, {H{\'e}brard}, {Udry}, {Bonfils}, {Delfosse},
  {D{\'e}sert}, {Sing}, \& {Vidal-Madjar}}]{2015Natur.522..459E}
{Ehrenreich}, D., {Bourrier}, V., {Wheatley}, P.~J., {et~al.} 2015, \nat, 522,
  459

\bibitem[{{Ehrenreich} {et~al.}(2008){Ehrenreich}, {Lecavelier Des Etangs},
  {H{\'e}brard}, {D{\'e}sert}, {Vidal-Madjar}, {McConnell}, {Parkinson},
  {Ballester}, \& {Ferlet}}]{2008A&A...483..933E}
{Ehrenreich}, D., {Lecavelier Des Etangs}, A., {H{\'e}brard}, G., {et~al.}
  2008, \aap, 483, 933

\bibitem[{{Foreman-Mackey} {et~al.}(2013){Foreman-Mackey}, {Hogg}, {Lang}, \&
  {Goodman}}]{2013PASP..125..306F}
{Foreman-Mackey}, D., {Hogg}, D.~W., {Lang}, D., \& {Goodman}, J. 2013, \pasp,
  125, 306

\bibitem[{{France} {et~al.}(2016){France}, {Fleming}, \&
  {Hoadley}}]{2016JATIS...2d1203F}
{France}, K., {Fleming}, B., \& {Hoadley}, K. 2016, Journal of Astronomical
  Telescopes, Instruments, and Systems, 2, 041203

\bibitem[{{France} {et~al.}(2017){France}, {Fleming}, {West}, {McCandliss},
  {Bolcar}, {Harris}, {Moustakas}, {O'Meara}, {Pascucci}, {Rigby},
  {Schiminovich}, \& {Tumlinson}}]{2017SPIE10397E..13F}
{France}, K., {Fleming}, B., {West}, G., {et~al.} 2017, in Society of
  Photo-Optical Instrumentation Engineers (SPIE) Conference Series, Vol. 10397,
  Society of Photo-Optical Instrumentation Engineers (SPIE) Conference Series,
  1039713

\bibitem[{{Gaudi} {et~al.}(2018){Gaudi}, {Seager}, {Mennesson}, {Kiessling},
  {Warfield}, {Habitable Exoplanet Observatory Science}, \& {Technology
  Definition Team}}]{2018NatAs...2..600G}
{Gaudi}, B.~S., {Seager}, S., {Mennesson}, B., {et~al.} 2018, Nature Astronomy,
  2, 600

\bibitem[{{Gillon} {et~al.}(2016){Gillon}, {Jehin}, {Lederer}, {Delrez}, {de
  Wit}, {Burdanov}, {Van Grootel}, {Burgasser}, {Triaud}, {Opitom}, {Demory},
  {Sahu}, {Bardalez Gagliuffi}, {Magain}, \& {Queloz}}]{2016Natur.533..221G}
{Gillon}, M., {Jehin}, E., {Lederer}, S.~M., {et~al.} 2016, \nat, 533, 221

\bibitem[{{Gillon} {et~al.}(2017){Gillon}, {Triaud}, {Demory}, {Jehin}, {Agol},
  {Deck}, {Lederer}, {de Wit}, {Burdanov}, {Ingalls}, {Bolmont}, {Leconte},
  {Raymond}, {Selsis}, {Turbet}, {Barkaoui}, {Burgasser}, {Burleigh}, {Carey},
  {Chaushev}, {Copperwheat}, {Delrez}, {Fernandes}, {Holdsworth}, {Kotze}, {Van
  Grootel}, {Almleaky}, {Benkhaldoun}, {Magain}, \&
  {Queloz}}]{2017Natur.542..456G}
{Gillon}, M., {Triaud}, A.~H.~M.~J., {Demory}, B.-O., {et~al.} 2017, \nat, 542,
  456

\bibitem[{{G\'omez de Castro} {et~al.}(2018){G\'omez de Castro},
  {Beitia-Antero}, \& {Ustamujic}}]{2018ExA...tmp....8C}
{G\'omez de Castro}, A.~I., {Beitia-Antero}, L., \& {Ustamujic}, S. 2018,
  Experimental Astronomy

\bibitem[{{Greene} {et~al.}(2016){Greene}, {Line}, {Montero}, {Fortney},
  {Lustig-Yaeger}, \& {Luther}}]{2016ApJ...817...17G}
{Greene}, T.~P., {Line}, M.~R., {Montero}, C., {et~al.} 2016, \apj, 817, 17

\bibitem[{Hunter(2007)}]{Hunter:2007}
Hunter, J.~D. 2007, Computing In Science \& Engineering, 9, 90

\bibitem[{Jones {et~al.}(2001)Jones, Oliphant, Peterson, {et~al.}}]{scipy_ref}
Jones, E., Oliphant, T., Peterson, P., {et~al.} 2001, {SciPy}: Open source
  scientific tools for {Python}

\bibitem[{{Jura}(2004)}]{2004ApJ...605L..65J}
{Jura}, M. 2004, \apjl, 605, L65

\bibitem[{{Kameda} {et~al.}(2017){Kameda}, {Ikezawa}, {Sato}, {Kuwabara},
  {Osada}, {Murakami}, {Yoshioka}, {Yoshikawa}, {Taguchi}, {Funase}, {Sugita},
  {Miyoshi}, \& {Fujimoto}}]{2017GeoRL..4411706K}
{Kameda}, S., {Ikezawa}, S., {Sato}, M., {et~al.} 2017, \grl, 44, 11

\bibitem[{{Kasting} {et~al.}(1993){Kasting}, {Whitmire}, \&
  {Reynolds}}]{1993Icar..101..108K}
{Kasting}, J.~F., {Whitmire}, D.~P., \& {Reynolds}, R.~T. 1993, \icarus, 101,
  108

\bibitem[{Kluyver {et~al.}(2016)Kluyver, Ragan-Kelley, P{\'e}rez, Granger,
  Bussonnier, Frederic, Kelley, Hamrick, Grout, Corlay, Ivanov, Avila, Abdalla,
  \& Willing}]{Kluyver:2016aa}
Kluyver, T., Ragan-Kelley, B., P{\'e}rez, F., {et~al.} 2016, in Positioning and
  Power in Academic Publishing: Players, Agents and Agendas, ed. F.~Loizides \&
  B.~Schmidt, IOS Press, 87 -- 90

\bibitem[{{Kopparapu} {et~al.}(2013){Kopparapu}, {Ramirez}, {Kasting}, {Eymet},
  {Robinson}, {Mahadevan}, {Terrien}, {Domagal-Goldman}, {Meadows}, \&
  {Deshpande}}]{2013ApJ...765..131K}
{Kopparapu}, R.~K., {Ramirez}, R., {Kasting}, J.~F., {et~al.} 2013, \apj, 765,
  131

\bibitem[{{Kreidberg}(2015)}]{2015PASP..127.1161K}
{Kreidberg}, L. 2015, \pasp, 127, 1161

\bibitem[{{Kulikov} {et~al.}(2007){Kulikov}, {Lammer}, {Lichtenegger}, {Penz},
  {Breuer}, {Spohn}, {Lundin}, \& {Biernat}}]{2007SSRv..129..207K}
{Kulikov}, Y.~N., {Lammer}, H., {Lichtenegger}, H.~I.~M., {et~al.} 2007, \ssr,
  129, 207

\bibitem[{{Kulow} {et~al.}(2014){Kulow}, {France}, {Linsky}, \&
  {Loyd}}]{2014ApJ...786..132K}
{Kulow}, J.~R., {France}, K., {Linsky}, J., \& {Loyd}, R.~O.~P. 2014, \apj,
  786, 132

\bibitem[{{Lammer} {et~al.}(2003){Lammer}, {Selsis}, {Ribas}, {Guinan},
  {Bauer}, \& {Weiss}}]{2003ApJ...598L.121L}
{Lammer}, H., {Selsis}, F., {Ribas}, I., {et~al.} 2003, \apjl, 598, L121

\bibitem[{{Lavie} {et~al.}(2017){Lavie}, {Ehrenreich}, {Bourrier}, {Lecavelier
  des Etangs}, {Vidal-Madjar}, {Delfosse}, {Gracia Berna}, {Heng}, {Thomas},
  {Udry}, \& {Wheatley}}]{2017A&A...605L...7L}
{Lavie}, B., {Ehrenreich}, D., {Bourrier}, V., {et~al.} 2017, \aap, 605, L7

\bibitem[{{Lecavelier Des Etangs}(2007)}]{2007A&A...461.1185L}
{Lecavelier Des Etangs}, A. 2007, \aap, 461, 1185

\bibitem[{{Lecavelier des Etangs} {et~al.}(2012){Lecavelier des Etangs},
  {Bourrier}, {Wheatley}, {Dupuy}, {Ehrenreich}, {Vidal-Madjar}, {H{\'e}brard},
  {Ballester}, {D{\'e}sert}, {Ferlet}, \& {Sing}}]{2012A&A...543L...4L}
{Lecavelier des Etangs}, A., {Bourrier}, V., {Wheatley}, P.~J., {et~al.} 2012,
  \aap, 543, L4

\bibitem[{{Lecavelier Des Etangs} {et~al.}(2010){Lecavelier Des Etangs},
  {Ehrenreich}, {Vidal-Madjar}, {Ballester}, {D{\'e}sert}, {Ferlet},
  {H{\'e}brard}, {Sing}, {Tchakoumegni}, \& {Udry}}]{2010A&A...514A..72L}
{Lecavelier Des Etangs}, A., {Ehrenreich}, D., {Vidal-Madjar}, A., {et~al.}
  2010, \aap, 514, A72

\bibitem[{{Lecavelier des Etangs} {et~al.}(2004){Lecavelier des Etangs},
  {Vidal-Madjar}, {McConnell}, \& {H{\'e}brard}}]{2004A&A...418L...1L}
{Lecavelier des Etangs}, A., {Vidal-Madjar}, A., {McConnell}, J.~C., \&
  {H{\'e}brard}, G. 2004, \aap, 418, L1

\bibitem[{{L{\'e}ger} {et~al.}(2004){L{\'e}ger}, {Selsis}, {Sotin}, {Guillot},
  {Despois}, {Mawet}, {Ollivier}, {Lab{\`e}que}, {Valette}, {Brachet},
  {Chazelas}, \& {Lammer}}]{2004Icar..169..499L}
{L{\'e}ger}, A., {Selsis}, F., {Sotin}, C., {et~al.} 2004, \icarus, 169, 499

\bibitem[{{Linsky} {et~al.}(2010){Linsky}, {Yang}, {France}, {Froning},
  {Green}, {Stocke}, \& {Osterman}}]{2010ApJ...717.1291L}
{Linsky}, J.~L., {Yang}, H., {France}, K., {et~al.} 2010, \apj, 717, 1291

\bibitem[{{Lothringer} {et~al.}(2018){Lothringer}, {Benneke}, {Crossfield},
  {Henry}, {Morley}, {Dragomir}, {Barman}, {Knutson}, {Kempton}, {Fortney},
  {McCullough}, \& {Howard}}]{2018AJ....155...66L}
{Lothringer}, J.~D., {Benneke}, B., {Crossfield}, I.~J.~M., {et~al.} 2018, \aj,
  155, 66

\bibitem[{{Lovis} {et~al.}(2017){Lovis}, {Snellen}, {Mouillet}, {Pepe},
  {Wildi}, {Astudillo-Defru}, {Beuzit}, {Bonfils}, {Cheetham}, {Conod},
  {Delfosse}, {Ehrenreich}, {Figueira}, {Forveille}, {Martins}, {Quanz},
  {Santos}, {Schmid}, {S{\'e}gransan}, \& {Udry}}]{2017A&A...599A..16L}
{Lovis}, C., {Snellen}, I., {Mouillet}, D., {et~al.} 2017, \aap, 599, A16

\bibitem[{{Ment} {et~al.}(2018){Ment}, {Dittmann}, {Astudillo-Defru},
  {Charbonneau}, {Irwin}, {Bonfils}, {Murgas}, {Almenara}, {Forveille}, {Agol},
  {Ballard}, {Berta-Thompson}, {Bouchy}, {Cloutier}, {Delfosse}, {Doyon},
  {Dressing}, {Esquerdo}, {Haywood}, {Kipping}, {Latham}, {Lovis}, {Newton},
  {Pepe}, {Rodriguez}, {Santos}, {Tan}, {Udry}, {Winters}, \&
  {W{\"u}nsche}}]{2018arXiv180800485M}
{Ment}, K., {Dittmann}, J.~A., {Astudillo-Defru}, N., {et~al.} 2018, ArXiv
  e-prints [\eprint[arXiv]{1808.00485}]

\bibitem[{{Muslimov} {et~al.}(2018){Muslimov}, {Bouret}, {Neiner}, {L{\'o}pez
  Ariste}, {Ferrari}, {Viv{\`e}s}, {Hugot}, {Grange}, {Lombardo}, {Lopes},
  {Costerate}, \& {Brachet}}]{2018arXiv180509067M}
{Muslimov}, E., {Bouret}, J.-C., {Neiner}, C., {et~al.} 2018, ArXiv e-prints
  [\eprint[arXiv]{1805.09067}]

\bibitem[{{Ochsenbein} {et~al.}(2000){Ochsenbein}, {Bauer}, \&
  {Marcout}}]{2000A&AS..143...23O}
{Ochsenbein}, F., {Bauer}, P., \& {Marcout}, J. 2000, \aaps, 143, 23

\bibitem[{{Oelkers} \& {Stassun}(2018)}]{2018arXiv180302316O}
{Oelkers}, R.~J. \& {Stassun}, K.~G. 2018, ArXiv e-prints
  [\eprint[arXiv]{1803.02316}]

\bibitem[{{Picone} {et~al.}(2002){Picone}, {Hedin}, {Drob}, \&
  {Aikin}}]{2002JGRA..107.1468P}
{Picone}, J.~M., {Hedin}, A.~E., {Drob}, D.~P., \& {Aikin}, A.~C. 2002, Journal
  of Geophysical Research (Space Physics), 107, 1468

\bibitem[{{Poppenhaeger} {et~al.}(2013){Poppenhaeger}, {Schmitt}, \&
  {Wolk}}]{2013ApJ...773...62P}
{Poppenhaeger}, K., {Schmitt}, J.~H.~M.~M., \& {Wolk}, S.~J. 2013, \apj, 773,
  62

\bibitem[{{Redfield} \& {Linsky}(2008)}]{2008ApJ...673..283R}
{Redfield}, S. \& {Linsky}, J.~L. 2008, \apj, 673, 283

\bibitem[{{Ricker} {et~al.}(2014){Ricker}, {Winn}, {Vanderspek}, {Latham},
  {Bakos}, {Bean}, {Berta-Thompson}, {Brown}, {Buchhave}, {Butler}, {Butler},
  {Chaplin}, {Charbonneau}, {Christensen-Dalsgaard}, {Clampin}, {Deming},
  {Doty}, {De Lee}, {Dressing}, {Dunham}, {Endl}, {Fressin}, {Ge}, {Henning},
  {Holman}, {Howard}, {Ida}, {Jenkins}, {Jernigan}, {Johnson}, {Kaltenegger},
  {Kawai}, {Kjeldsen}, {Laughlin}, {Levine}, {Lin}, {Lissauer}, {MacQueen},
  {Marcy}, {McCullough}, {Morton}, {Narita}, {Paegert}, {Palle}, {Pepe},
  {Pepper}, {Quirrenbach}, {Rinehart}, {Sasselov}, {Sato}, {Seager},
  {Sozzetti}, {Stassun}, {Sullivan}, {Szentgyorgyi}, {Torres}, {Udry}, \&
  {Villasenor}}]{2014SPIE.9143E..20R}
{Ricker}, G.~R., {Winn}, J.~N., {Vanderspek}, R., {et~al.} 2014, in \procspie,
  Vol. 9143, Space Telescopes and Instrumentation 2014: Optical, Infrared, and
  Millimeter Wave, 914320

\bibitem[{{Rogers}(2015)}]{2015ApJ...801...41R}
{Rogers}, L.~A. 2015, \apj, 801, 41

\bibitem[{{Sachkov} {et~al.}(2016){Sachkov}, {Shustov}, \& {G{\'o}mez de
  Castro}}]{2016SPIE.9905E..04S}
{Sachkov}, M., {Shustov}, B., \& {G{\'o}mez de Castro}, A.~I. 2016, in
  \procspie, Vol. 9905, Space Telescopes and Instrumentation 2016: Ultraviolet
  to Gamma Ray, 990504

\bibitem[{{Selsis} {et~al.}(2007){Selsis}, {Chazelas}, {Bord{\'e}}, {Ollivier},
  {Brachet}, {Decaudin}, {Bouchy}, {Ehrenreich}, {Grie{\ss}meier}, {Lammer},
  {Sotin}, {Grasset}, {Moutou}, {Barge}, {Deleuil}, {Mawet}, {Despois},
  {Kasting}, \& {L{\'e}ger}}]{2007Icar..191..453S}
{Selsis}, F., {Chazelas}, B., {Bord{\'e}}, P., {et~al.} 2007, \icarus, 191, 453

\bibitem[{{Shizgal} \& {Arkos}(1996)}]{1996RvGeo..34..483S}
{Shizgal}, B.~D. \& {Arkos}, G.~G. 1996, Reviews of Geophysics, 34, 483

\bibitem[{{Snellen} {et~al.}(2015){Snellen}, {de Kok}, {Birkby}, {Brandl},
  {Brogi}, {Keller}, {Kenworthy}, {Schwarz}, \& {Stuik}}]{2015A&A...576A..59S}
{Snellen}, I., {de Kok}, R., {Birkby}, J.~L., {et~al.} 2015, \aap, 576, A59

\bibitem[{{Snellen} {et~al.}(2013){Snellen}, {de Kok}, {le Poole}, {Brogi}, \&
  {Birkby}}]{2013ApJ...764..182S}
{Snellen}, I.~A.~G., {de Kok}, R.~J., {le Poole}, R., {Brogi}, M., \& {Birkby},
  J. 2013, \apj, 764, 182

\bibitem[{{Sullivan} {et~al.}(2015){Sullivan}, {Winn}, {Berta-Thompson},
  {Charbonneau}, {Deming}, {Dressing}, {Latham}, {Levine}, {McCullough},
  {Morton}, {Ricker}, {Vanderspek}, \& {Woods}}]{2015ApJ...809...77S}
{Sullivan}, P.~W., {Winn}, J.~N., {Berta-Thompson}, Z.~K., {et~al.} 2015, \apj,
  809, 77

\bibitem[{{Vidal-Madjar} {et~al.}(2003){Vidal-Madjar}, {Lecavelier des Etangs},
  {D{\'e}sert}, {Ballester}, {Ferlet}, {H{\'e}brard}, \&
  {Mayor}}]{2003Natur.422..143V}
{Vidal-Madjar}, A., {Lecavelier des Etangs}, A., {D{\'e}sert}, J.-M., {et~al.}
  2003, \nat, 422, 143

\bibitem[{{Volkov} {et~al.}(2011){Volkov}, {Johnson}, {Tucker}, \&
  {Erwin}}]{2011ApJ...729L..24V}
{Volkov}, A.~N., {Johnson}, R.~E., {Tucker}, O.~J., \& {Erwin}, J.~T. 2011,
  \apjl, 729, L24

\bibitem[{{Wallace} {et~al.}(1970){Wallace}, {Barth}, {Pearce}, {Kelly},
  {Anderson}, \& {Fastie}}]{1970JGR....75.3769W}
{Wallace}, L., {Barth}, C.~A., {Pearce}, J.~B., {et~al.} 1970, \jgr, 75, 3769

\bibitem[{{Wenger} {et~al.}(2000){Wenger}, {Ochsenbein}, {Egret}, {Dubois},
  {Bonnarel}, {Borde}, {Genova}, {Jasniewicz}, {Lalo{\"e}}, {Lesteven}, \&
  {Monier}}]{2000A&AS..143....9W}
{Wenger}, M., {Ochsenbein}, F., {Egret}, D., {et~al.} 2000, \aaps, 143, 9

\bibitem[{{Wheatley} {et~al.}(2017){Wheatley}, {Louden}, {Bourrier},
  {Ehrenreich}, \& {Gillon}}]{2017MNRAS.465L..74W}
{Wheatley}, P.~J., {Louden}, T., {Bourrier}, V., {Ehrenreich}, D., \& {Gillon},
  M. 2017, \mnras, 465, L74

\bibitem[{{Wilson} {et~al.}(2017){Wilson}, {Lecavelier des Etangs},
  {Vidal-Madjar}, {Bourrier}, {H{\'e}brard}, {Kiefer}, {Beust}, {Ferlet}, \&
  {Lagrange}}]{2017A&A...599A..75W}
{Wilson}, P.~A., {Lecavelier des Etangs}, A., {Vidal-Madjar}, A., {et~al.}
  2017, \aap, 599, A75

\bibitem[{{Wood} {et~al.}(2005){Wood}, {Redfield}, {Linsky}, {M{\"u}ller}, \&
  {Zank}}]{2005ApJS..159..118W}
{Wood}, B.~E., {Redfield}, S., {Linsky}, J.~L., {M{\"u}ller}, H.-R., \& {Zank},
  G.~P. 2005, \apjs, 159, 118

\bibitem[{{Youngblood} {et~al.}(2017){Youngblood}, {France}, {Loyd}, {Brown},
  {Mason}, {Schneider}, {Tilley}, {Berta-Thompson}, {Buccino}, {Froning},
  {Hawley}, {Linsky}, {Mauas}, {Redfield}, {Kowalski}, {Miguel}, {Newton},
  {Rugheimer}, {Segura}, {Roberge}, \& {Vieytes}}]{2017ApJ...843...31Y}
{Youngblood}, A., {France}, K., {Loyd}, R.~O.~P., {et~al.} 2017, \apj, 843, 31

\bibitem[{{Youngblood} {et~al.}(2016){Youngblood}, {France}, {Parke Loyd},
  {Linsky}, {Redfield}, {Schneider}, {Wood}, {Brown}, {Froning}, {Miguel},
  {Rugheimer}, \& {Walkowicz}}]{2016ApJ...824..101Y}
{Youngblood}, A., {France}, K., {Parke Loyd}, R.~O., {et~al.} 2016, \apj, 824,
  101

\end{thebibliography}

\end{document}